# 增强现实系统对在中国农村小学英语学习者的影响：动机, 成就, 行为和认知的实现

**IJAZ UL HAQ**

# 艾哈

**2017 年 06 月**

中图分类号：TQ028.1

UDC 分类号：540

# 增强现实系统对在中国农村小学英语学习者的影响:动机, 成就, 行为和认知的实现

| | |
|---|---|
| 作 者 姓 名 | 艾哈 |
| 学 院 名 称 | 计算机学院 |
| 指 导 教 师 | 牛振东教授 |
| 答辩委员会主席 | 刘辉教授 |
| 申 请 学 位 | 硕士学位 |
| 学 科 专 业 | 计算机科学与技术 |
| 学位授予单位 | 北京理工大学 |
| 论文答辩日期 | 2017 年 06 月 |

# Impact of Augmented reality system on elementary school ESL learners in country side of china: Motivations, achievements, behaviors and cognitive attainment

| | |
|---|---|
| Candidate Name: | IJAZ UL HAQ |
| Faculty Mentor: | Professor Dr. Zhendong Niu |
| Chair, Thesis Committee: | Professor Liu Hui |
| Degree Applied: | Masters of Science |
| Major: | Computer Science and Technology |
| Department: | School of Computer Science and Technology |
| Degree By: | Beijing Institute of Technology |
| Date: | June, 2017 |

增强现实系统对在中国农村小学英语学习者的影响:动机,成就,行为和认知的实现

北京理工大学

# 研究成果声明

本人郑重声明：所提交的学位论文是我本人在指导教师的指导下进行的研究工作获得的研究成果。尽我所知，文中除特别标注和致谢的地方外，学位论文中不包含其他人已经发表或撰写过的研究成果，也不包含为获得北京理工大学或其它教育机构的学位或证书所使用过的材料。与我一同工作的合作者对此研究工作所做的任何贡献均已在学位论文中作了明确的说明并表示了谢意。特此申明。

  特此申明。

  签　名：　　　　　　　　日期：

# 关于学位论文使用权的说明

本人完全了解北京理工大学有保管、使用学位文的定，其中包括：①学校有保管、并向有部送交学位文的原件与印件；②学校可以采用影印、印或其它制手段制并保存学位文；③学校可允学位文被或借；④学校可以学交流目的,制送和交学位文；⑤学校可以公布学位文的全部或部分内容（保密学位文在解密后遵守此定）。

  签　名：　　　　　　　　日期：

  导师签名：　　　　　　　日期：

# To my parents,

"My Lord! Have mercy upon them as they nourished me when I was small."

*Al-Quran*

"Paradise lies under the feet of the mother."

*Prophet Muhammad* (peace be upon him)

# 摘要

由于传统教学方式的落后，先进技术资源的缺乏和学习动机不强，农村和偏远地区学生的英语水平普遍低于城市地区的学生，这导致农村地区的学生英语成绩很差，学习效果不好。在这项研究中，我们通过使用基于增强现实的英语单词学习系统（AREWL）教授农村和偏远地区的小学生学习英语单词，研究了他们的学习动机，学习成绩，行为模式和认知水平。我们分析了是否因为成绩不同而导致学生学习动机不同，或者情况恰恰相反，同时也分析了学生的行为模式，认知水平以及二者之间的联系，揭示了学生与应用程序是如何交互的。AREWL 使用 3D 虚拟现实技术、动画和有趣的评估教授学生使用英语认读和拼写一些单词，为了方便学生使用，该手机软件的说明语言设置为英文和中文。本研究的样本组将依据学生的英语水平来选择。我们通过前测选择了一、二年级共 20 名小学生作为样本组，这些学生的前测成绩都很差。同时样本组里还包括五位非母语英语教师。关于数据收集，我们使用了前测和后测，问卷调查表，视频录像以及 AREWL 自带的学习评估系统。关于数据分析，我们使用了三角测量法。定量的方法将用来分析前测后测、教师意见，学习动机和学习成绩之间的关系。而定性方法将用来揭示学生的行为（通过摄像机录制）及其与认知水平的关系、学生的意见等。结果表明，教师和学生都很喜欢使用 AREWL，学生不管内在还是外在动机都被调动了起来。研究显示，他们的学习动机与学习成绩有显著的关系。而行为和认知水平方面的分析显示，学生与 AREWL 交互良好，但其认知水平并不高。从这些分析中我们得出结论，AREWL 对于中国年轻学习者的英语学习动机是很重要的。此外，本研究结果通过分析 AREWL 这款英语单词学习软件填补了教育技术领域的空白，为英语教育的发展做出了贡献。

关键词：增强现实，英语学习，动机，认知素养，行为模式。

# Abstract

The English level of students from countryside areas of china is lower than the students living in urban areas, because of use of old traditional ways of teaching, lack of advance technology resources and lack of English learning motivation, which causes poor outcome in students grades and language learning achievements. In this study learning motivation, learning achievement, behavioral patterns and cognitive attainment of the students are examined, by using augmented reality English words learning (AREWL) system for teaching English words to elementary schools students in countryside area of china. Within this context, it is discussed whether the motivation of student differs according to their level of achievements or vice versa, and also the analyses of behavioral patterns, cognitive attainment and their relation is revealed that how the students interact with the AREWL . 3d virtual objects, animations and fun assessments are used to teach students how to pronounce and spell name of these objects in English, the instructions of the applications are both in English and Chinese for the ease of students. The sample group of the study was selected on the basis of their English level, 20 students from grade 1 and grade 2 of elementary school are selected on the basis of pretest results score, the students with poor scores were selected, and the sample also included 5 non-native teachers. Pretest and posttest, questioners, survey forms, video recordings, and within the application evaluation of student learning achievements are used for collecting the data. Quantitative methods are used to analyze pretest-posttest, and teacher's opinion, and relations between motivation and achievement while qualitative methods are used to reveal the behaviors of students, recorded by the cameras and its relation with cognitive attainment, and also the opinions of the students.

The results shows that both teachers and students liked the AREWL activity, also it was revealed that the students were both intrinsically and extrinsically motivated, and their motivations had significant relations with their learning achievement, Behavioral, and cognitive attainment analysis shows that these students played interactively with the AREWL but cognitive attainment was not so high. From all these analysis we conclude that AR based learning applications are important for the motivation of English learning in china for young learners. Moreover the findings of this study provide important contribution to the education of this era by presenting a new AR based English words learning application to fill the gap in the field of educational technologies.

# CONTENTS

# List of Tables

# List of Figures

# 1 INTRODUCTION

## 1.1 BACKGROUND OF THE STUDY

From the Past few decades education of English language is considered as the subject of importance in the mainland of China, and English language proficiency has been regarded widely as a personal and as well as national asset [1], [2], [3],[4] Chinese leadership view English as a vital role player in the development and modernization of the society [5],[6]). Although English was firstly taught in China in 1862, after the foundation of PRC, the cultural revolution of Chinese became a block to the English education development, so until 1970's it did not become an essential part of education in China. The status of English learning started changing in 1980's when English teaching was included in the syllabus of elementary schools to university. [7], [8]. However, for many decades the teaching and learning of English have been highly ignored by both Western and Chinese scholars. With the advancement of time, the situation for English learning is changing, the urban areas of China have good schools and universities, who use advance way of teaching but rural areas or country side people are facing problem in the way of getting admission in these high-level universities because of the gap of education between urban and rural areas students. [9]. The three major reasons are lack of use of advance educational resources, bilingual policies of education and limited use of English language.

There is an important role of educational resources in every level of education. PRC made some significant improvements, but still, there is no much satisfaction, specifically in the rural and country side of China. Shortage of books, educational facilities, and poor teaching staff, have created a gap in rural and urban education [10], according to researches, there is no difficulty in finding extremely poor schools in country sides of China [11]. [12] mentioned, large numbers of schools in rural areas have a lack of advance educational resources for improving their listening and speaking abilities. Although the number of resources increased in recent years, still there is a shortcoming in advance technological resources for English teaching. [13] stated that some of the teachers are not even willing to do jobs in the rural area's schools because of unsatisfactory conditions of schools and career plans and some of them want to quit their jobs. One example from the literature is about schools in Sichuan Province, the education for grade4 to grade 6 w not provided, so all the students of that area had to join the primary school in the area of Baiwu Town to continue their education which was quite far to reach for many students [14].

Looking at these circumstances the leadership of China is taking steps forward to provide cheap education to those areas which are not so advanced. In this study, we are focusing on providing a cheap way of learning English for elementary schools students by using Augmented Reality technology which will motivate student's English learning, and they can learn at home and both at schools. Because it is known that lack of motivation causes lack of interest in learning English, if knowledge is presented interestingy, students feel happy to take part in the learning process, in this study we will examine learner's motivation by introducing a new technology into their English learning process and analyzed how they behave while using this technology for learning, also their cognitive attainment and achievements after the use of Augmented Reality English words learning application.

## 1.2 PROBLEM STATEMENT

The English level of students from rural or countryside areas of China is lower than the students living in urban areas [15], because of use of old traditional ways of teaching, lack of use of advance technology, and not having good teaching staff. Although in the last four decades social advancement and remarkable economic growth has been achieved by China but still there exist a huge gap between rural and urban areas [16]. The population of China is 1.3 billion among which 800 million live in rural areas, and the earning of those people is coming mainly from agriculture. China took various steps to promote development in country side zone s. From 2004 to 2009 various measured s were taken by the Chinese government, for the implementation of nine years of basic education [16]. Teachers were encouraged for working in rural areas to improve the quality of teaching and by using new ways of learning [16].

Despite of all these efforts done by the government the schools of country side areas in China are still undeveloped because of old traditional ways of teaching, teaching quality and lack of use of technologies, which causes low motivation in English learners, because in schools they are mostly focusing on writing and reading but have fewer chances for speaking. So people from rural areas face this problem of speaking when they need to adapt themselves to urban communities, where communication becomes a problem and low motivation is caused which effect the learning achievements [16]. The government also have adopted many e-learning methods for English language learning; the low achievement ha caused low motivation [17].

These problems can be addressed by making the use of cheaper technologies common, nowadays mobile phone have many learning applications available for learning English, Augmented Reality technology that is new widely available, easy to use and less expensive, the applications of AR are attractive, because it mixes the real and virtual objects together to make the learning process more appealing.

## 1.3 AIMS AND OBJECTIVE OF THE RESEARCH

In this study we introduce a new teaching environment by emphasizing on the values of both intrinsic and extrinsic motivation of English learners in rural areas of China by using Augmented Reality English words learning AREWL System, to explore the effectiveness of this teaching methodology in terms of student motivation, learning achievement, behavioral patterns, cognitive attainment and the relation between English learning motivation of students and the learning achievements further more relation between behavioral patterns and cognitive attainment during learning will be analyzed by using AR English words learning system.

So with in this context, we will answer the following research questions.

Q1. What are the achievements of students using AREWL?

Q2. Were there both intrinsic and extrinsic motivations involved in the learning process?

Q3. What is the relation of student's intrinsic and extrinsic motivation with their learning achievement?

Q4. How do students behave while using AREWL?

Q5. What was the cognitive attainment of students using AREWL?

Q6. What is the relation between behavioral patterns and cognitive attainment of students?

Q7. What are the opinions of teacher and students about the use of AREWL?

## 1.4 RESEARCH CONTRIBUTIONS

1. The aim of this study is to investigate the effectiveness of using AR apps as a teaching and learning tool while instructing the elementary school students in rural areas of China.
2. Introduction to a new learning environment where the values of intrinsic and extrinsic motivation are emphasized equally.
3. Questions regarding learning motivations, achievement, behaviors, cognitive attainment, and the relation between them are answered
4. As the study regarding the use of AR apps is the new and hot topic, and especially there is low awareness of the use of this technology in the rural areas of China, so that i i is why a study of the effectiveness of AR technology in education is beneficial to the administrators and educators.

This kind of research can help to improve the education quality and can give encouragement for the adoption of new teaching methods.

## 1.5 SIGNIFICANCE OF THE RESEARCH

The contribution of new technologies to the educational fields are valuable, that is why researchers are trying to explore their effects on students learning, and their focus is on "How can modern software technologies support ubiquitous effective knowledge and learning management solutions" [18]. An AR applications are employed which can involve both teacher and the student in the learning process. [17] in his doctoral thesis suggested that such environments for language learning should be created where both intrinsic and extrinsic values should be emphasized equally, so in this study we developed an English learning system for increasing intrinsic motivation of student, and also a role has given to the teacher to evaluate the student learning by using the same system with teacher preferences, which will create extrinsic motivation.

[17] also found out that students having both intrinsic and extrinsic motivation towards English learning show more achievement in their grades as compared to those, having just extrinsic or

intrinsic motivation. In this regard the exploration of the relation between motivation and achievement in learning English using AR application is important.

However determining how the student interact with our AR system and to know which behavioral acts of students are important, because these relates to deeper examination in interaction with AR so for this Analysis of student behavioral patterns will be performed.

## 1.6 DEFINITION OF TERMS

Some of the terms used in this study need to be interpreted in this study for the readers, so we provide the following lists of definitions.

*Liangxiang County*: this county is situated in the countryside of Beijing, China

*Technology*: there are many definitions of technology, but for the purpose of our study, we are referring to, mobile phone, computers, LCD, Smart Boards, and Projectors, etc.

*Augmented Reality*: This is a new technology which uses the camera to the augmented 3d object in a real world environment.

*Elementary School*: schools which have classes from grade 1 to grade 5.

*Elementary Teacher*: the teachers who teach in elementary school

*Intrinsic Motivation*: if the student is motivated by some personal interest in doing something without the effect of surroundings.

*Extrinsic Motivation*: If some external rewards cause the motivation, and the external cause effects the motivation is called extrinsic motivation.

*Behavioral Patterns*: in this study, we refer behaviors patterns as those behaviors of individuals which the do while using some technology for knowledge gain.

*Cognitive Attainment*: it is the term which refers to a mental process of knowledge gain and comprehension, it include knowing, thinking, remembering, problem-solving and judging.

## 1.7 ASSUMPTIONS

We have made several assumptions in this study; the first assumption is the participants involved in this study will answer all the questions honestly, as far as the matter of privacy, maybe some teachers have got this fear that their answers will reach the higher administrative authorities, so they may not answer honestly. The second assumption of this study is that participant of this study will be similar to other rural parts of China as well.

## 1.8 SCOPE AND LIMITATIONS

The study took place in a small private school of Lianxiang County, we wanted every students should have one smartphone, which should be provided by the family or the school, but some students could not bring because of some problems so we provided them, and internet was also needed for using this application, so sometime internet had problems, some participant may have just skipped the learning part quickly to see the other content of the application and they did not attempt all the steps in the application. While filling survey forms the participants (student, and teachers), the students may have answer some questions quickly because of lack of understanding and teachers because of having the administrator's fear, have answered some questions dishonestly.

Another Limitation of the study is that we did sampling by purposive way, the selection of school in liangxiang was based on the condition and resources of the school, other schools in that area maybe not bad and are facilitated, secondly the student sample was based on the result of pretest, so maybe the poor cause of students result based on pretest is that technology use was not common and student despite of having good knowledge attempt the pretest poorly, so these factors create a chance to reduce generalizability of the findings.

## 1.9 SUMMARY

Chapter 1 introduce the problems for English learners in Country side areas of china, and explained that many factors are included which results the poor English knowledge gain in undeveloped part of the country, most of the problems discussed was related to demographic allocations, lack of use of technologies less educational resources, poor teaching staff, etc of country side schools.

Further we highlighted the problem statement, and propose possible solution for the problems, also there is discussion on the contribution of the study, and discussion on the questions to be answered.

In the next chapter title as Literature Review, we will put light on the previous researches done in that area and will reveal the gap in those researchers.

# 2 REVIEW OF LITERATURE

## 2.1 DEFINITION OF MOTIVATION

In Psychology, the term "Motivation" is well known and is one of the main topics in education. The word motivation is derived from "*movere*" which is a Latin word, whose meaning is to move [19]. If someone is motivated, it means he/she is moved to doing something.

They also explained that "the idea of movement is reflected in such commonsense ideas about motivation as something that gets us going, keeps us moving and helps us get jobs done."

[20] defined motivation in terms of psychology as "motivation is commonly thought of as an inner drive, impulse, emotion or desire that moves one to a particular action." Another definition is "an innate state that arouses, directs and maintains behavior" [21]. According to psychologist motivation is a "drive" which for the satisfaction of ones need, give a direction to invest their energies.

[22] in the educational point of view defined motivation as "motivation is something that energizes, directs and sustains behavior; it gets students moving, points them in a particular direction, and keeps them going." For motivation of someone it is important to increase the interest, maintain that interest and spending time and energy for the achievement of goals [23]. From this, it can be observed that motivation involves goals. [24] stated that

*"The motivated individual expends effort, is persistent and attentive to the task at hand, has goals, desires, and aspirations, enjoys the activity, experiences reinforcement from success and disappointment from failure, makes attributions concerning success and/ or failure, is aroused, and makes use of strategies to aid in achieving goals".*

[21] made a conclusion about motivation as "the study of motivation is essentially a study of how and why people initiate actions directed towards specific goals and persist in their attempts to reach these goals."

In comparison to achievement, the researchers mostly focus on the two types of motivation, Intrinsic and extrinsic motivation. [19] define these terms as

*"Intrinsic motivation refers to motivation to engage in an activity for its sake. People who are intrinsically motivated work on tasks because they find them enjoyable. Task participation is its reward and does not depend on explicit rewards or other external constraints. In contrast, extrinsic motivation is motivation to engage in an activity as a means to an end. Individuals who are extrinsically motivated work on tasks because they believe that participation will result in desirable outcomes such as a reward, teacher praise or avoidance of punishment."*

Both of these types of motivations are essential for good learning outcomes, we will further discuss these types of motivation in next section, but at this point, it is also important to consider about motivation in foreign language learning.

## 2.2 ROLE OF MOTIVATION IN FOREIGN LANGUAGE LEARNING

Motivation has the main role in learning a foreign language, if a student is not motivated enough, his/her achievements are mostly low. [25] mentioned that "it has chiefly been used to refer to the long-term relatively stable attitudes in the students' minds." Another researcher [26] pointed out that there are three elements for the identification of foreign language learning motivation, stated as

1. A wish to learn language
2. Effort used for learning the language
3. Positive behavior towards learning the language

So we can simply say that motivation is a process in which to achieve the long-term commitment, a student's wish, his/her persistence and the positive attitude towards the goal of language learning, during the process of learning is considered. [27] stated that "motivation determines the extent of active, personal involvement in L2 learning". [20] point out that "countless studies and experiments in human learning have shown that motivation is a key to learning."In addition [28] mentioned that motivation ascertains the level of the learner's deep and active involvement and learning attitude. [29] describes the role of language learning motivation in their famous socio-psychological model regarding integrative and instrumental motivation as

*"The orientation is said to be instrumental in the form of the purposes of language study reflect the more utilitarian value of linguistic achievements, such as getting ahead in one's occupation. In contrast, the orientation is integrative if the student wishes to learn more about the other cultural community because he is interested in it in an open-minded way, to the point of eventually being accepted as a member of that other group."*

It seems that motivation has important role in the language learning, different researchers defined motivation from different point of views, the reason behind this is because language learning has different context. However, the main point to notice is the key to learning a language is motivation.

In one place [22] describes motivation in learning point of view as "students who wants to master themselves in a specific course material are more seems to be intrinsically motivated as compared to those who wants to perform well in a course have chances that they are extrinsically motivated. In other words, we can say that those students who have the desire of achieving some goal in their learning should be intrinsically and extrinsically motivated, in this relation these two types of motivations are discussed in the next section.

## 2.3 RELATION BETWEEN MOTIVATION AND ACHIEVEMENT

The results of many previous types of research show that motivational factors are related for sure to learning of a foreign or second language [30, 31].It means that for successfully learning a language motivation is one of the key factors, the reason behind this is the dependency of achievement on learner's effort used in a foreign language learning task.

From another point of view, achievements in learning a language can help to increase student's motivation. [32] mentioned about achievement motivation that it depends on the wish to succeed, which means the tendency to struggle for success or excellence. [33] point out the relation between the motivation and achievement as "the need to attain a standard of excellence, or to accomplish a goal to prove one's worth." When people have the ambition to achieve, then they work hard and have a high attitude for making sure that they are successful. If they have intrinsic motivation, then they will take part in an activity for the sake of improvement or learning, but If they have the extrinsic motivation, then they take part in the activity with the expectancy of rewards. Perseverance and willingness in language learning are linked closely to their needs of achievement.

[34] says that in the motivational factors of learning a foreign language the achievement needs plays a major role. The reason behind this is because learning a foreign language mostly takes place in academic contexts or schools that can be described as an achievement in academics situations. Therefore, the need of the individual for achievement will significantly affect learning.

1n another point of view, the acquisition contexts of the second language can relatively provide many different ways for the attainment of language, for example communicating with native speakers of the language. Therefore, the lack of achievement needs to be compensated by a strong motivation.

'Resultative hypothesis' of [35] exemplify that motivation may be the result of some achievements in the learning experience. In other words, those students who perform well are motivated or reinforced to work hard as compared to those who do not perform so good are discouraged by their low performance which causes demotivation. [36] in his research tests the resultative hypothesis. The findings of her work show "the motivational hypothesis does not fully account for the interrelationship between attitudes and success or failure in the second-language acquisition process. They also reveal that foreign-language learning causes the formation of positive and negative attitudes" [36], which means that the reason behind someone's learning motivation is a success or we can say that success in learning results motivation. [36] also mentioned that those learners who perform well develop more motivational intensity and likely to be more active in the classroom.

From the evidence above we bring out the issue that whether the learners are motivated because of their achievement or the reason behind their motivation is the success in learning of language [37] stated that "the relationship between motivation and achievement is an interactive one." In other words, we can say that there is a contribution of the high level of motivation in the learning process and also achievement can help to enhance or maintain the existing motivation, as compared to this Conversely, if the motivation is low which will cause low achievement that further will cause lower motivation in learning.

The research findings suggest that there is an effective relationship between achievement and motivation in learning a foreign language [38].In simple words, the high in the motivation the more is a success, the more the success, the higher is the motivation. The discussion of this dependency is important, [36] stated that

"It could be tremendously important for a teacher to know whether the motivational or the resultative component is predominant at particular stages of the language course, and to adjust teaching procedures and strategies accordingly."

From the above evidence we can conclude that success enhances motivation and motivation enhance the success, but the problems lie in the vicious circle negative mirror, which means lower achievement cause lower success, and vice versa.

[39] stated that "a vicious circle of low motivation → low achievement → lower motivation can develop especially if learners attribute their failure to factors they feel powerless to alter."

In short we can say that motivation and achievements are closely related to each other, in current study we will focus on the low motivation caused among Chinese students because of low interest in English language learning, we will discuss this reason in the coming section, but at this stage we will discuss the effect of intrinsic motivation on achievement and also the effect of extrinsic motivation on leaning achievement. So in the next subsection, we will discuss it.

## 2.4 THE IMPACT OF INTRINSIC AND EXTRINSIC MOTIVATION ON ACHIEVEMENT

Researchers have been working in this area since last few decades. Motivation learning is classified into intrinsic and extrinsic motivation. In 1980's self-determination theory was introduced in which main focus was on motivational behavior that leads to achievements whether the motivation is extrinsic or intrinsic. [40] in this literature importance of intrinsic motivation was also discussed as "being intrinsically motivated to learn improves the quality of learning and that those conditions that are autonomy supporting and informational will promote more efficient learning as well as enhanced intrinsic motivation and self-esteem." Intrinsic motivation work and learning is one that is interesting, and these activities are performed by students to do for themselves to gain joy, satisfaction, and fun [41]. In 1993 [41] in his work describes Intrinsic motivation as a deep approach learning whereas Extrinsic motivation as a surface approach learning. Intrinsic motivation is developed in an individual or student because of his desire because it is done for his sake of doing and on the other hand, extrinsic motivation is developed due to the outward factors that could be for the sake of getting benefits or achieving high grade or any other reward.

Many researchers have compared the results and findings of extrinsic and intrinsic motivation, [42] explained that intrinsic motivation has a great importance as compared to extrinsic motivation. In his work, it was also discussed that when a student works and learns something for his own then intrinsic motivation reasons i.e. attaining interesting knowledge, better understanding and development of skills related to learning causes pleasure and enjoyment to them. While on the other side it was investigated that if the reasons behind motivation are extrinsic that are not for the sake of themselves but the sake of others one of the main factors of extrinsic motivation could be fear of getting punished or getting bad grades and not getting all the benefits and reward. There is no self-desire force present in extrinsic motivation like it was present in intrinsic motivation. Intrinsic motivation caused a strong and positive change in student learning as they are more persistent learners.

[40] mentioned that there is a chance that tendency towards low performance to lack the ability or intelligence can result if the learner's orientation is extrinsic if this is the case the learner will find himself helpless and the motivation chances for further improvement will be less. On the other hand, if the learner's orientation is intrinsic learners may be more tend to attribute a low performance to an unstable factor or variable like effort and thus accept it as a challenge which will motivate him to try harder. This kind of contrastive responses can describe that why the learners who are intrinsically motivated may learn better. The reason behind this is that because the learners having intrinsic motivation may be less likely to accept the feelings of low self-and esteem and helplessness when they see failure. Behaviors which shows extrinsic motivation are considered to be less self-determined and possessed by external constraints as compared to those who are intrinsically motivated.

When the main concern of a learner is to get an extrinsic goal, then s/he will try to complete the task as early and quickly and with as little frustration and effort as possible [43]. This type of learning approach has low chances to lead to effective learning. [34] Claims that learner who has intrinsic motivation is more likely to gain a higher language proficiency than a learner who is extrinsically motivated. Therefore we can say that intrinsically motivated students are more likely to satisfy themselves for the achievement of a goal.

On the other hand, it can be argued that both extrinsic and intrinsic motivations be necessary for schools environment. However, of course, the ideal thing is to create an intrinsically motivated environment which may satisfy the needs of students. However, it is sometimes difficult for teachers to create intrinsic motivation among students all the time as not everything in the classroom is not so interesting which can give intrinsic motivation. Moreover, learning about each student personally take long period. As a result, students may sometimes have the need to know about the significance of extrinsic motives and for engaging themselves in learning they rely on these reasons. As an example, students commonly need to prepare for examinations and study. Extrinsic motivation has undeniable role in the curriculum of the schools. [40] stated that it is unlikely that extrinsic and intrinsic motivation will be exclusively experienced by a learner. In some situations external rewards are important.

From all the points mentioned above, we can say that teachers should pay attention to extrinsic and intrinsic motives of the student so that students can have practical experiences and enjoy learning English.

According to the author of [44], both extrinsic and intrinsic motivations have a major role in the classroom environment because teachers influence for accessing them is least partially. Sometimes external incentives are involved in the derivation of extrinsic motivation. Such as peer group or parent influences, which is inaccessible to the influence of a teacher. However, there are some other reasons of extrinsic motivation that a teacher can affect. For example, the teacher the teacher use some rewards to reinforce behaviors of desired students. Similarly, teacher can discourage the student's undesirable behavior if h/she use some punishment, so external incentive of this kind may help to motivate students and which in a classroom always exists.

On the other hand, [44] in his research work argues that the roots of intrinsic motivation belong to previous attitudes of the student. This means if students experience while learning is worthwhile or not. Bad experiences of some English learners, mostly develop a hostile attitude towards learning English.

From all the evidence above, it is clear that both intrinsic and extrinsic motivation have a strong relation with students learning achievements. Moreover, the most important among both of them is the creation of intrinsic motivation among students.

As the current research focus on the motivation and achievements among English learners in the rural or country side area of China, so in the next section, we will discuss the motivation and the achievement level of Chinese learners.

## 2.5 ENGLISH LEARNING MOTIVATION IN CHINESE STUDENTS

In the past ten years, China has increasingly integrated into the global community, culture, economy, and the establishment of a high-profile presence in various fields at the international stage.

This emphasizes the importance of learning English in the country, [45] mentioned that in 2005 about 17.7 million people are engaged in some English learning, which is the world's largest English learning group in the world. Despite this boom, there have been sporadic reports on the motivational basis of language learning from this important country. English proficiency in China is seen as an asset in the individual and fairly clear Social level in this era of economic globalization. Therefore, English is taught as a core subject from elementary till the second year of the university. However, unified educational goal is not achieved in a uniform environment of learning. As the population of China is above one billion, due to which it shows considerable regional differences. [46] mentioned that In the past 30 years, the economic growth of the eastern region of China is rapid, the development opportunity is obvious, and the economic growth is faster in the Western and central regions, also [47] say that the increment in the regional development gap is still there. Regional economic disparities, resulting in uneven distribution of educational resources in these areas, [48] Therefore, it is not surprising that students in more developed areas of the country are found to have more proficiency in English than their peers in any part of China [49].

We find a further contrast in development point of you within the region. That is the gap between rural and urban areas. The gap between rural and urban areas in China is further widening because of the scarceness of social and natural resources. The impact of this on the education of English language is inevitable because urban areas are usually assigned better teachers, teaching, curriculum resources and equipment [50]. As a result of the recognition of heterogeneity, a large number of students in rural areas are sent to schools in urban areas to receive a better education in urban schools, although often quite expensive [51].

In addition to the urban-rural and regional differences, we find another major difference in the education system of Chinese. In higher education there exist two tiers: ordinary universities and key university, which was launched by the government to promote the "211 Project" "985 Project" (Ministry of education of the people's Republic of China 2009) [48].The level of English proficiency varies not just because of regional or geographic aspects but also because of differences in the universities.

As there are many factors which affect the learner's motivation in English learning[52] in his research discuss age related differences in the motivation of English learners in China.

From the results of his work, he found that age of the student is related to his/her motivational development, and English learning motivation shifts at a different level of schooling from elementary to high school. The above evidence suggest that instrumental motivation among students starts when they enter in middle schools. The elementary students begin to learn English with impartial motives.

The interrogative motivation among them is higher than secondary school peers but similar to the peers of middle school as compared to the instrumental motivation among them which is mostly similar to both secondary and middle school peers. [52]

## 2.6 AGE-RELATED MOTIVATIONAL DIFFERENCES AMONG ENGLISH LEARNERS IN CHINA

Although the majority of students learning English in Mainland of China are children and most of what is known about motivation among of the learners of English language in China is limited to the studies of the adults and high school students. The advantage of learning a second language young children are more than adults [53].The children in the ages from 3-6 already know the purpose and the way to use a language. Moreover, if children at the early age are exposed to learn two languages, they have the ability to learn both and keep both of the languages separate, and also can have the understanding of using the language with different timings, people and places appropriately [54]. Previous researches show that adults or older children are an initially better learner of English, but younger children in the long run are better in the context of second language acquisition (SLA). Eventually young learners always surpass and catch up the adults or older children [55] ,However there is a difference in the age impact of learning a foreign language as compared to the second language, ESL students are likely to learn more efficiently because of the environment in which he lives belongs to the target language, on the other hand in the environment of EFL the availability of language is limited to the formal instructions in schools or learning centers, where the interaction with target language are provided few hours a week by non-natives, so chances of interaction and using the foreign language outside the class is low. (Cenoz, 2003). Therefore in comparison of ESL environment with EFL environment, the young learners in EFL environment need longer time to learn or catch up. [55].

In the current study we focus on the children or students of elementary school living in rural or country side of China, to find some ways to motivate these students with cheap and easily available technology, because nowadays embedding technology in education is a hot topic, so in the next section we will discuss, the use of technology in education, its significance and the research done in that area.

## 2.7 THE USE OF TECHNOLOGY IN EDUCATION

Technology plays an important role for improvement of education system it is a daunting task to divide the effects of other factors from the effect of technology which has impact on learning and teaching [56]. The use of technology into the process of learning means that there is a need of innovation, teachers to become Innovative in the classroom is a lengthy process that will take years to wait [57]. [58] Stated that "As students begin to develop technology habits, it is vital to teach them how to effectively use the tools available to them." Technology is considered to be "a body of knowledge and actions, used by people, to extend the human potential for controlling and modifying the natural and human-made environment" [56]. Integration of technology does not happen by chance it seems educational plan created with a purpose that requires knowledge and vocational guidance [59]. The integration of technology should simple, and it should be an integral part of modern education in the classroom [59]. The resources of technology include dedicated software, computer systems and network communications and other equipment. [60]. [59] Maintains that "Technology should not only be seen as a writing tool, similar to a pen, pencil, or paper-lined notebook."

[61] in his research mentioned about the learners of the twenty-first century that their level of engaging themselves in education and learning is much higher and they require fast access to every new information and are very relational. [61] stated that we cannot say in this modern era that student has less technological resources than teachers. Digital artwork nowadays can be done independently by this era learners, and it has become easy for them after the implementing the standards of International Society for Technology (ISTE) [61] The use of modern software, has made it easy for young learners to write and edit their movies and story books. [61]. [62] stated that if the talent of students in the class is recognized by the teacher, and teacher finds that the students can use the technology independently, and have the ability to design some good digital work. So the next step of the teacher is to concern about fitting all the expected work into a school day. [61] mentioned that the teacher expectations of using technology in the classroom were the teacher-centered type teaching by using interactive technology in the form of multimedia presentation to the students $21^{st}$ Century.

The above evidence show that there are countless benefits of technology in the field of education. In the next subsection, we will discuss the benefits of Technology in Education.

## 2.8 BENEFITS OF THE USE OF TECHNOLOGY IN EDUCATION

This is the belief of researchers that if the technology is integrated into education, it leads to success and achievements [63].Collaborating through social networks will give students the potential to create knowledge, share content, and create new online communities in the cloud.

[62] explained that students collaboration with the help of social networks would give them the potential to create new online communities and knowledge. [63] Stated that there is an endless potential for learning if the technological devices are in the hands of students. The need of the time is that teachers should be a helper of students in the development of their technological skills, which can further be integrated for improving other areas in a subject, which will result in positive learning outcome [64].

[65] Mentioned that the students studying in the district's school and who are disabled nowadays have more access and help available which provide help to the students in solving many issues. If the classrooms are more advanced and the school's administrators provide more technological

ways of teaching, it will result in the increment of student's ability to access and save more information [66]. Back in time teachers used to tell their students how to learn and what they should learn, and gave them the definition of knowledge. However, now students have the ability and also the resources which make them learn by their own [67]. Administrators of schools should consider it as one of the cores duty to encourage technological use to support new educational learning tools which will support the learning of the students. [68].

[69] in her article mentioned few benefits of technology in education some of them we will summarize here, she mentioned that technology have countless benefits if used for education, it makes the teaching easy, because teachers are full of technological resources, presentations, audio visuals and video aids like thing have made teaching interesting and easy for them. It is no harder to track the student study records, and the progress he/she makes, are under the eyes of teachers. The use of technology in education is also good for the environment, like if someone imagines, that back in time how many paper and pencils and trees were used to fulfill the student's requirement, now teacher can make his or her own online test which saves many resources. Few decades before student were not finding education as interesting and it seemed boring for them but thanks to the technology tools which have made learning fun and interesting for students. The Internet nowadays has minimized distances; distance learning is not a problem now, the teacher can teach student from any part of the world.

In simple, we can conclude that it is difficult to imagine education without the use of modern technologies. Many technologies are available, and researchers have been working to use the latest technologies in education fields, E- learning is not just limited to learning online from the internet, nowadays mobile learning have made the learning process easier, especially the emergence of new technology like Augmented Reality have to change the level of leaning which provides a new way of interaction, it allows the user to interact with real and physical surrounding [70]. The use of this new technology in the educational field is under focus by the researchers. In the next section, we will put light on some importance and the use of augmented reality AR.

## 2.9 AR DEFINITION

The definition of Augmented Reality (AR) differs among researchers of Computer Science and education. According to [71] AR combines virtual and real objects which provides the opportunity for real-time interaction and give accurate registration of three-dimensional virtual and real object. [72] definition of augmented reality is "a situation in which a real word context is dynamically overlaid with coherent location or context-sensitive virtual information." [73] mentioned that "Augmented reality (AR) is an emerging technology with the potential to transform learning. By digitally adding or removing information from the physical world, AR creates a sense that real and virtual objects coexist, and can enhance people's interactions both with each other and with objects in the world."

In the current study we are using the technology of augmented reality technology to teach English words to students of age 5-7, in rural or country side of China to find out its effectiveness or, As AR technology is not expensive and it provides interesting way of learning which is enjoyable by students, and they are motivated to learn more, in the upcoming section we will discuss the effectiveness, use, and benefits of Augmented reality in education and then specifically for learning English.

In the upcoming section, we will discuss the work done by other researchers who used AR in education, will discuss if there are any benefits of using it, and then specifically will discuss the work done for learning English using this technology.

## 2.10 THE USE OF AUGMENTED REALITY IN EDUCATION

After the emergence of AR technology, the researchers started to explore its uses and check the possibility where it can be more useful, there is a lot of work done on it in different educational field,

The table (1) below will show the Researches done on AR in various educational fields

Table 1: Use of AR in different educational fields

| Educational Field | References |
|---|---|
| Medical education | [74], [75], [76], [77] |
| Physics Education | [78],[79] ,[80] |
| Museum Education | [81], [82], [83], [84] |
| Biology Education | [85],[86], [87] |
| Astronomy Education | [88],[89], [90] |
| Chemistry Education | [90], [86],[91] |
| Geometry & Mathematics Education | [92],[93],[94] |

## 2.11 THE BENEFITS OF AUGMENTED REALITY EDUCATIONAL APPLICATIONS

There are numerous benefits of augmented reality in learning, if used in a purposeful direction, the applications of augmented reality can enhance the attention of the learner, and he/she can focus for more time on the learning material [80]. It means if the student pays more attention to any learning material, he/she will learn and understand it more. [95],[96] mentioned that educational applications of augmented reality technology could provide the effective experience of learning because it is attractive for the users. The reason for the attraction in this technology lies in its mixing the real world object with the virtual objects. Many researchers find out that augmented reality AR based learning application help in enhancing the learner's motivation [97],[80]. Augmented reality applications are interactive, user have to interact with the 3d objects which become the part of the real worlds, so in one way we can say that this technology enhances the interaction between of the user with the content of the application [98],[99], [100], [101]. This technology facilitates the learning process, and keep the learner engaged for long time. So the more the learner is involved in the learning activity, the better he/she learns [102], [100],[101]. With the passage of time, the use of augmented reality applications in learning is increasing day by day because of its numerous benefits. The applications of augmented reality AR are cooperative, as it makes the learning process easy, and triggers the creativity of the learners, and make him able to think differently [103], [104], [105].The way in which the augmented reality applications works are amazing, as it plays the role of developing the imagination of the learner [105], all these benefits of Augmented Reality AR make this technology perfect for creating learning environments, in which the learners have the self-interest to study. [96] stated that this technology enhances the spatial thinking of the user. Almost every in every field of education, importance is given to this emerging technology to improve the quality of education, and to motivate students for self-study, English learning is also nowadays one of its area in which researcher are working on in its different aspects. The current study focuses on its use in teaching English to Chinese kids belonging to rural or urban areas, to provide them the cheap and attractive way of English learning. In the next section, we will discuss, about the approaches used to teach English using augmented reality AR.

## 2.12 AUGMENTED REALITY FOR LEARNING ENGLISH

In the last few decades many researchers have put efforts to make Augmented Reality AR useful for teaching and learning.Few of them also work on English learning using AR, [106] implemented an English vocabulary learning AR system to investigate how much learning satisfaction is gained after the learning process, what are the behavioral intentions of students towards English vocabulary learning using AR and to find out how much effective the learning process can be. The results show that few factors including, quality of the system and operation of the process affected the perceived usefulness & perceived satisfaction, the suggested that the learning process should be straight forward, so that they can be easy for use by the users.

[107] worked on AR for teaching English to Malaysian primary school students to find out if the there is any increment on the motivation towards learning English in students? The followed the Keller's ARCS motivational model, the results of the study were positive, there was a significance change in the motivation of students.

Also it is stated that those who learn English using AR applications, they acquire more comprehension, reading, speaking & listening skills as compared to the English learner of the traditional way [108],[109],[109].

[110] examined attitude, achievement, and cognitive load levels of the students learning English using AR, the digitized one lesson of Grade 5 English curriculum book, and developed an AR

English Learning system, the results of the study show positivity in regard to students attitude, and relation between achievement and cognitive load levels.

In another study in Kuwait [110] conducted research on kindergarten students, teaching them English alphabets using AR, his finding shows that kids like learning alphabets with an interactive way of learning, the results shows that the learning process was effective. Table (2) shows the researches which have been done in the area of English learning using Augmented Reality Technology.

## 2.13 TOOLS USED FOR DEVELOPING AR APPLICATIONS

Although there are many tools available for developing AR applications such as Wikitude (http://www.wikitude.com), Layar (http://www.layar.com), and Vuforia (www.vuforia.com). In the current study, we choose Vuforia because it is easy to use and the online community is active. Moreover its tracking stability is excellent and also much help is available for the developers.

## 2.14 SUMMARY

The study above is related to the effort made till now, to improve the quality of English learning techniques used for the motivation of Chinese students to learn English with interest. Many researchers have mentioned that motivation has a vital role in ones learning. Specifically the role of students achievement with respect to his/her motivation is the important matter of concern to be addressed, the students who live in the country side part, are not much aware that English can be learned in many interesting ways with the help of technology, Augmented reality is one of among that modern technology for motivating student to learn, because it have the ability to gain student's attention for longer time, the work related to English learning by the use of augmented reality have been in limelight since a long time, researchers from different countries are doing research to find the impact of this technology on different sample groups. From kindergarten to university level teaching. In most of the study research questions are answered by doing an analysis of the different pattern of learning.

Our work is different from the previous researchers because we have tried to find the answer to several questions which were not answered before and also we have tried to respond to some questions similar to other researchers but we did the experiment on the different sample than those. Firstly we found out from the study, that for gaining more achievements regarding students learning outcomes, both intrinsic and extrinsic motivation are necessary for high achievement. So researchers suggested that such environments are important for effective results of English learning in which value of both intrinsic and extrinsic motivation are equally emphasized so, for this our English words learning model have reward based section which is related to extrinsic motivation and an augmented reality-based learning which cause intrinsic motivation because the student is eager to see the magic of augmented reality and try to pass the lesson evaluation to see other magical learning material in the next day lesson. Secondly we also examined the behavioral pattern and cognitive attainment of students from rural parts of China, which previously were done for the different experiment of learning through augmented reality technology but not specifically for learning English.

# 3 DEVELOPMENT OF AUGMENTED REALITY ENGLISH WORDS LEARNING APP (AREWL)

## 3.1 DESIGN AND DEVELOPMENT OF THE APPLICATION

The system developed in this study is to check the effectiveness of AR on Chinese elementary schools students for basic English words learning, this application involves both teacher and the student in the learning process. The application includes, 3d models, animation, and audio visuals, which are presented with the integration of AR technology, which will make learning as a fun activity for non-motivated students, the application focuses on learning simple English words, like the name of things, etc. As the sample group is from China, so for making it easy for use, both English and Chinese instructions are used. The student will learn one lesson daily; the student can unlock the lesson of next day if he/she can pass the daily lesson assessment, a lesson has 3 to 4 words to teach, for example, fruits, animal, vehicle, numbers, etc. . Figure 1 shows the flow chart diagram of the whole learning Process of AREWL. It shows the two portals being developed, one is for teacher and one is for students, the flow chart shows the relation between student and teacher portal and also the role of both of the portals. As this is a demo application for checking the effectiveness of AR technology in terms of students learning motivation, so the quality of graphics and interface is not so good, some off the graphics were developed by taking help of 3d designers and some of the graphics are taken from internet which are freely available for use.

### 3.1.1 Development tools

The Application is developed in Unity 3d which is a game development engine, we imported Vuforia AR development Kit to Unity 3D for development of Interactive learning portal, MY SQL server is used in the backend for storing student's and teacher's data. 3D Max software is used for development of 3D objects. Unity 3d uses 2 scripting languages, JavaScript and C#, in this application scripting is done using C#.

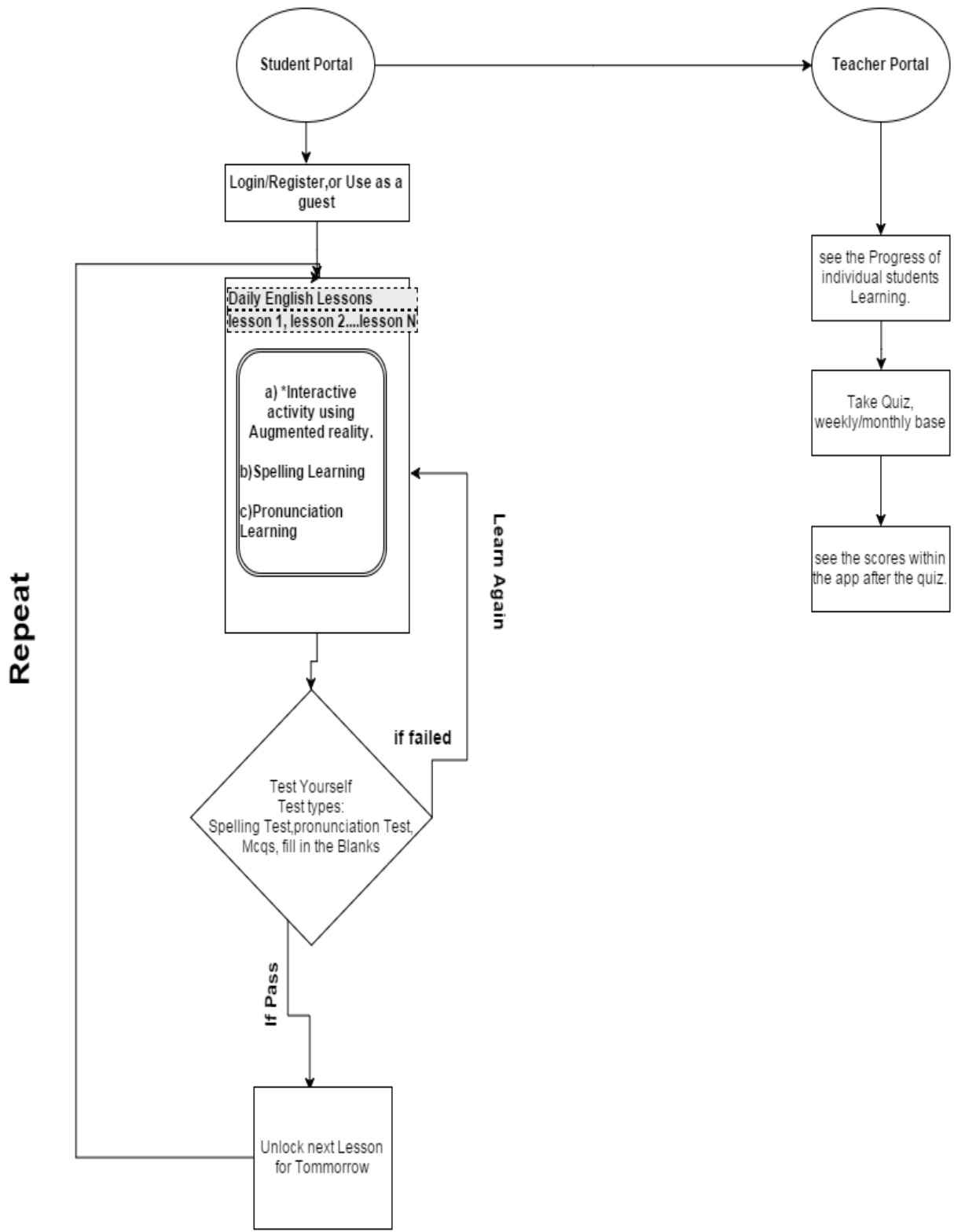

Figure 1: Flow Chart Diagram of the AREWL

There are two portals of this application, student portal and teacher portal Figure 2

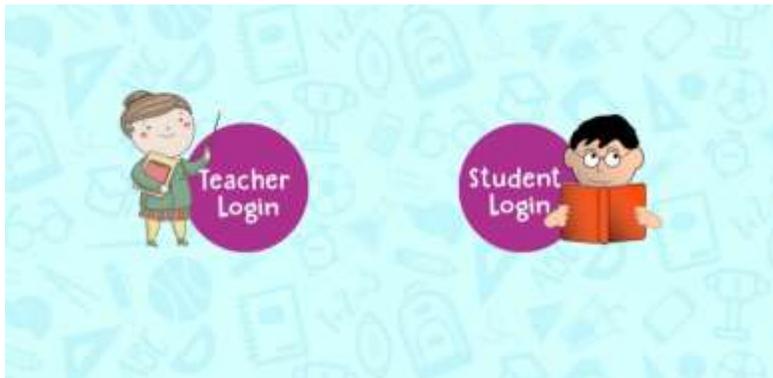

Figure 2: Teacher and Student Login

## 3.2 Student Portal

Students have to learn one lesson daily lesson, the content of lesson were developed according to a book used for teaching English words, so we in one way digitized it and with the use of technology made it more attractive for the students. The lesson which is active for the current day have a green color, and the other two lessons are disable shown in blue color. Figure (5)

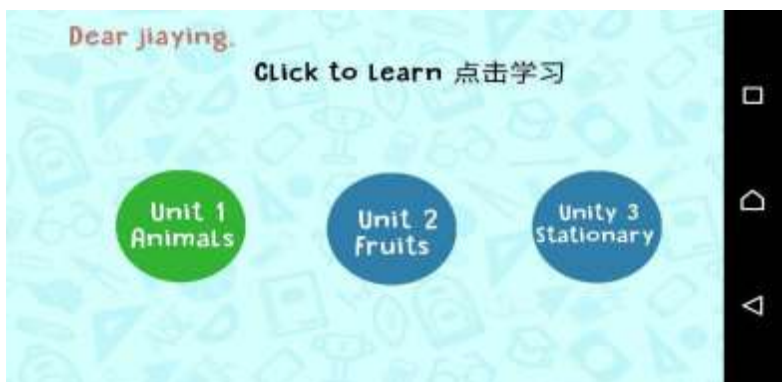

Figure 5: Daily Learning Lesson screen

Every unit has some learning materials, figure (6) shows an example of it. In unit 1 four animals are taught, cat, wolf horse and tiger, each of them is taught interactivey, to learn what the animal is and how to write it  English, and listen to learn pronunciation. So learning every animal name is a must thing to do for unlocking the learning process of another animal.

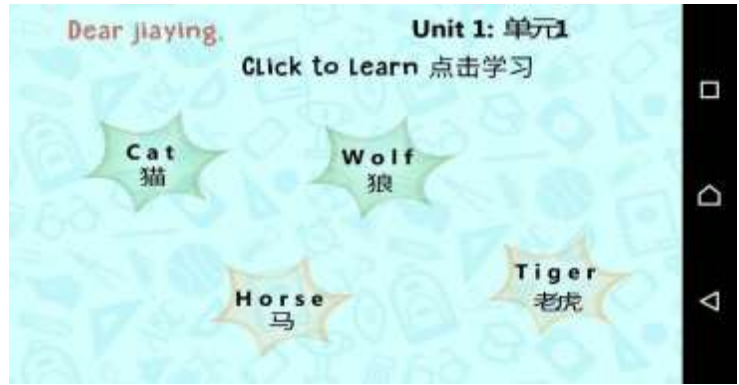
Figure 6: Demo Lesson 1 Screen

This application have different learning stages, we have used mix methods of teaching, by this we mean that along with augmented reality, there are other contents for teaching, e.g. for spelling learning and assessment we have a 2D interface within the application, following are the learning processes of the application.

1. Interactive learning
2. Spelling Learning
3. Pronunciation Learning
4. Learning assessment

### 3.2.1 Interactive Learning

In this step the student will use the book or flash cards with a smartphone or a tablet, when the learner see any specific picture for learning within the application, the application will convert it into an interactive activity using augmented reality technology, for example, if the first lesson is about learning name of animals, so those animals mentioned on the book page will be converted into interactive 3d object on the top of book page which will put some fun in learning, and the student will remember it easily the next time he/she see it. Within this learning, the student can interact with the 3d object, can touch it, animate it, and can listen to audios related to its animation and name pronunciation. Figure (7) shows the sample picture of augmented reality interactive learning process.

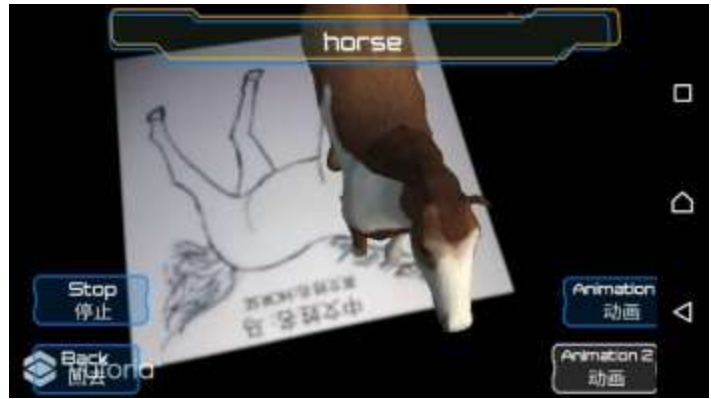
Figure 7: AR Interactive Learning Screen

### 3.2.2 Spelling Learning

This part will focus on learning the spellings of things by letting the student practice on learning the spelling of things. In this section, the student is thought how to write the spelling of a word. First the spelling of the object is appeared one by one along with the pronunciation. Then after student ha to type the same spelling as taught. If the student types the spelling wrong, then he/she have to type again, unless and until they type it correctly. Figure (8) shows an example of spelling learning process.

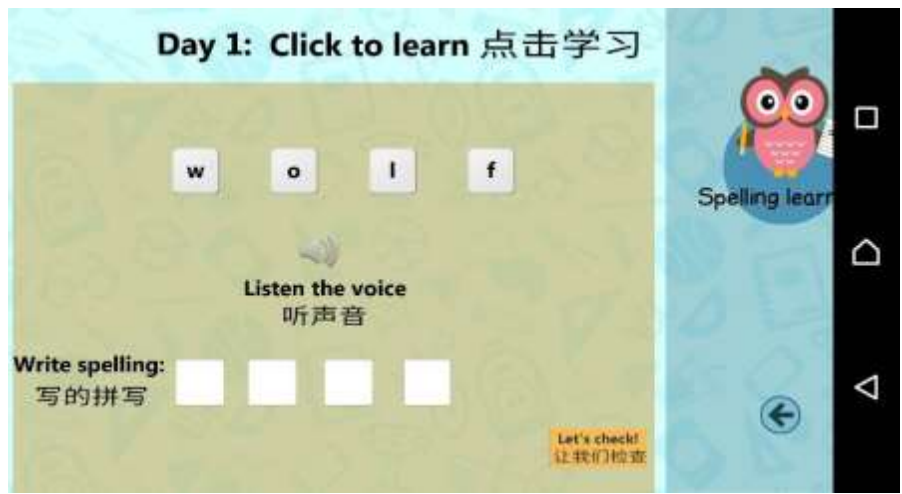
Figure 8: AREWL Spelling Learning Screen

### 3.2.3 Pronunciation Learning

This part will mainly focus on the learning of pronunciation of a word in proper accent, so every new word have its pronunciation available in the system, the student have to listen and learn how to pronounce, the listening part is not a separate part of the learning process, we assume that a student while playing with interactive learning and spelling learning part, at the same time he/she also listen to the word pronunciation and is becoming familiar with the pronunciation of the that word.

### 3.2.4 Daily Learning Assessment

After the learning process of each daily lesson, the assessment of each lesson is unlocked, in which few simple questions related to that daily lesson are asked, to check if the student can recognize the object, can spell the name of that object, and do he/she know the pronunciation. Figure 9 shows the type of learning assessment used in the application, if the student make mistakes, in attempting the questions of assessment, then he/she can have a as many attempt as he/she wants until he complete it without any mistake, if the assessment is done once without making any errors in the answer then the student can unlock the next lesson for tomorrow, this locking and unlocking of the lesson is because, it's a fun to learn through this application, and the students will find it interesting to explore every object, using AR technology, which will be interesting for them and enjoyable for them, so we assumed that for unlocking the next lesson for the next day the student have to learn the previous lesson without mistakes, which also give the surety of learning. Figure (9), (10), (11) and (12).

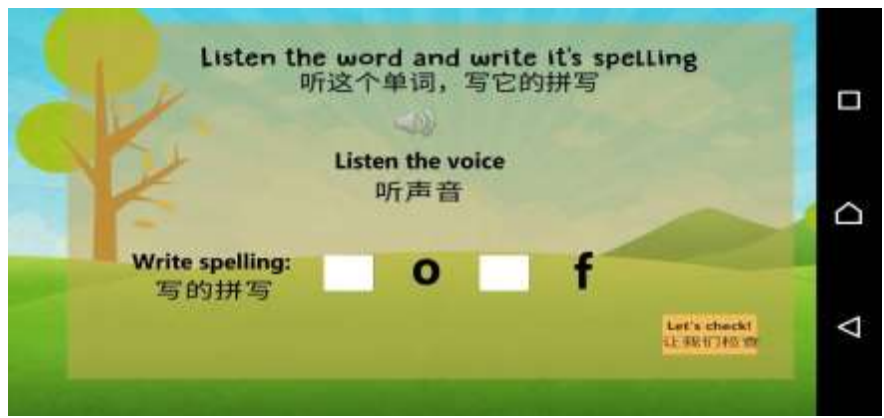

Figure 9: Daily lesson assessment screen 1

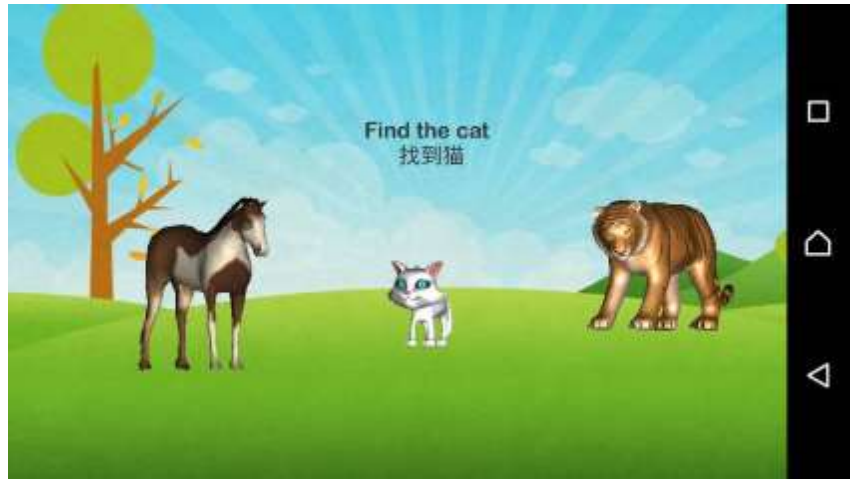
Figure 10: Daily lesson assessment screen 2

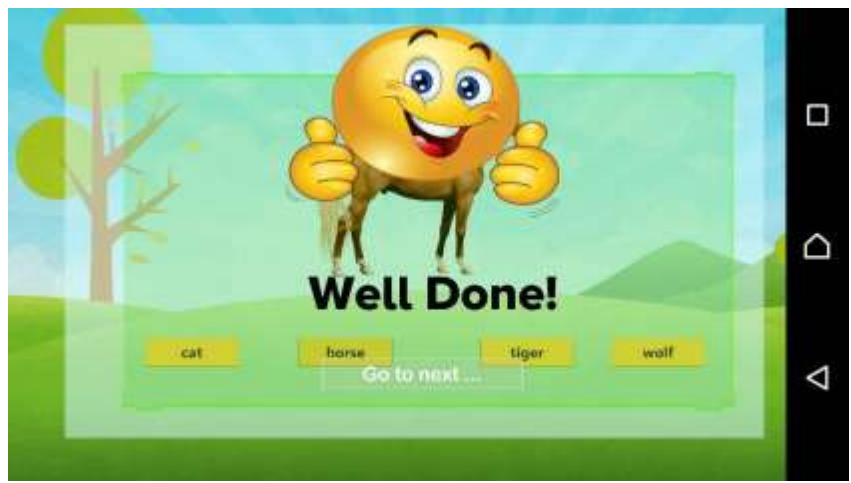
Figure 11: Daily lesson assessment screen 3

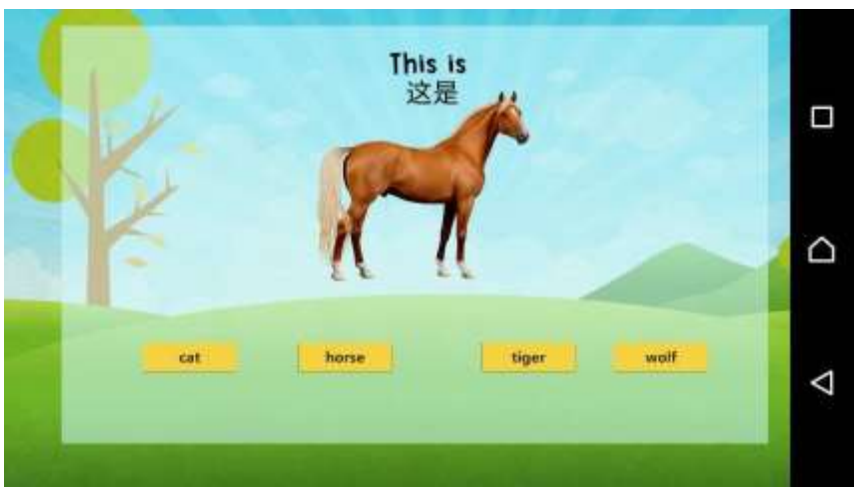
Figure 12: Daily lesson assessment screen 4

## 3.3 TEACHER PORTAL

The role of the teacher is important in any learning activity, the teacher should know the learning progress of his/her students, so that he/she can pay attention to those students who are a weak learner, in this English learning application we have developed a teacher interface, teacher have the following privileges.

1. Register a student
2. See the learning progress of individual student (attempts and mistake)
3. Take a quiz

### 3.3.1  Register a student

In this part, a student can register a new student, who become the part of teacher's group. For this study, we registered five students with one teacher.

### 3.3.2 Learning Progress of student

As the student will be learning one lesson per day, and there is a chance that he/she would have done some mistakes and also had re-attempt the assessment many times until the assessment is passed without any mistakes, so a teacher can see each student, attempts, and mistakes of a specific assessment, and know the learning level of each student. Figure (13) and (14) shows an example of student's attempts and mistakes records, along with teacher name and the students register with him/her.

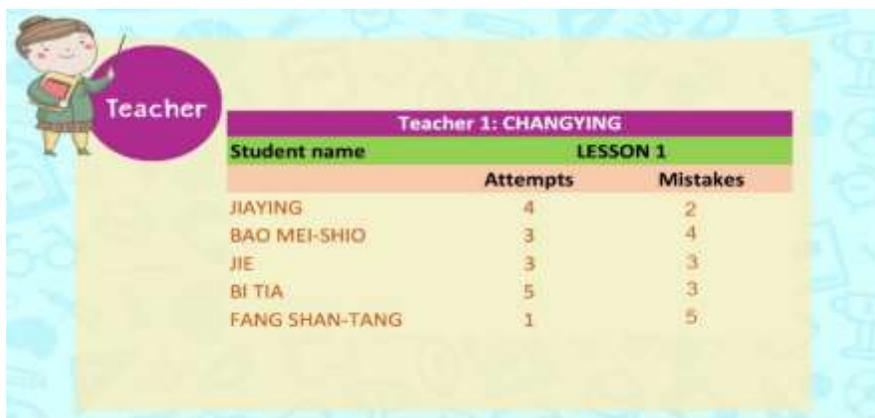

Figure 13: Daily Lesson Attempts and mistakes

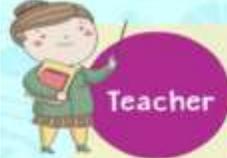

Figure 14: Daily Lesson Attempts and mistakes

### 3.3.3 Take a quiz

Teacher have this option that he/she can take a quiz from all those students together who are registered with that specific teacher. The quiz can be among any number of lesson learned, the main purpose of this quiz is to extrinsically motivate the student to compete for getting good scores, because, it is kind of reward based. After the students attempt the quiz, the teacher portal does an evaluation of each student result and display a list of student's scores to the teacher, from high to low score. Which means the student whose score is high is on the top of the list similarly the one have ha the lowest is at the bottom of the list. Figure (15).

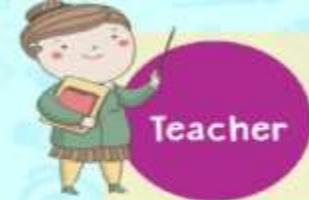

Figure 15: Weekly Test Score of 20 Students

## 3.4 COMPARISON OF AREWL WITH OTHER AR BASED ENGLISH LEARNING APPLICATIONS

AREWL is different from other AR based English learning application, in many ways, in different regions the AR for English learning has been tested on various sample groups, and the researches wanted to find the relations between the various variables differently a y. In our work, we focused on Elementary Schools students who had very poor English knowledge because of low motivational reasons in country side of Beijing.

Following are the major differences between our application and work from the others.

1. The region of the experiment is different as, we chosen sample from the rural area of China.
2. AREWL focus on teaching not only using AR, like the other applications, just have interaction in the learning process.
3. Three steps are learning, interactive learning for recognition of a new object, learn how to spell, and pronounce, while others have different approaches.
4. Daily evaluation ensures the learning of a specific lesson, which previous works do not have.
5. Relation with the teacher is also one of the main difference which was included for extrinsic motivational factors; other applications just have augmented reality part.
6. Quiz by a teacher in all the learning activities, including the quiz in interactive learning part.
7. Fun learning part is AR based learning which will create intrinsic motivation in students, and Combine quiz between all the students, create extrinsic motivation, which improves the results,

As compared to other learning AR application developed for similar researches which mainly focus, either alphabets learning, or just vocabulary, AREWL combines some features of the previous works, to make a proper English fun learning activity.

Table 2: Comparison of AREWL with previous Researches

| Author's | Country/Region | Title of the study | Main Findings | AREWL |
|---|---|---|---|---|
| [111] | Taiwan, China | QR Code and Augmented Reality-Supported Mobile English Learning System | Location based mobile learning by decrypting the QR code at different locations, the system was proved to be effective for learning. | **Region:** China (Country side). 1.Teacher and student Relation in the application, 2. Teaching focus on interaction, spelling & pronunciation learning 3. Test Your Self 4. Teacher preferences: Register a student, monitor learning progress, and take a quiz. 5. Intrinsic and extrinsic motivations creation during the activity. 6. Finding relations between motivations and achievements, |
| [112] | Japan | Applying Augmented Reality to E-Learning for Foreign Language Study and its Evaluation. | Augmented reality based learning was found to be less stressful than traditional English learning | |
| [113] | Kuwait | The Effectiveness of Using Augmented Reality Apps in Teaching the English Alphabet to Kindergarten Children: A Case Study in the State of Kuwait | A significance difference in learning outcomes of the experimental group, the group which used AR for learning got more results in the evaluation. | |
| [110] | Turkey | Augmented Reality for Learning English: Achievement, | The levels of a Cognitive load of AR based | |

|   |   | Attitude and Cognitive Load Levels of Students | environment are low. Successful students attitude were higher. | and also behaviors and cognitions. |
| --- | --- | --- | --- | --- |
| [106] | Taiwan, China | Investigating students' perceived satisfaction, behavioral intention, and effectiveness of English learning using augmented reality | Behavioral intentions of students were affected by perceived usefulness and perceived satisfaction. |   |
| [107] | Malaysia | The Use of Augmented Reality Pop-Up Book to Increase Motivation in English Language Learning For National Primary School | Used ARCS Keller's model to find motivations of the learners, the learners were motivated. |   |

## 3.5 SUMMARY

In this chapter, we discussed the AREWL application which had been developed as a demo application for current research. As we are concerned about the increasing student's motivation and to find out if there is a relation between students learning motivation and the achievements regarding scores? In one way we want to test our learning process algorithm which we think can create a learning environment which will motivate the students intrinsically by having fun learning the part, and extrinsically by competing for the reward to become the best learner of the week. As we performed our experiment in country side area, where schools are not much facilitated, so our point of interest was also here to find the way of interaction these students do, while using AREWL and also the cognitive attainment of students. In the coming section, we will discuss the whole experiment that how it was performed, and what were the finding of our study.

# 4 RESEARCH APPROACHES FOR GATHERING DATA

## 4.1 INTRODUCTION

We developed AREWL to test it on country side area elementary school students, the next step was to test it by doing an experiment on some group of students of a specific age, and at the same time, and we needed some data for answering our research questions. The selection of right data collection tools was necessary after a thorough study of literature we found out which tools have to be used.

We used triangulation methods, which include both Quantitative and qualitative analysis, Teachers opinion, student's intrinsic and extrinsic motivation, achievements, and behavioral patterns were analyzed by quantitative methods and Students opinion, and cognitive attainments were analyzed by qualitative methods. The sections below will show the whole research process step by step

## 4.2 THE PURPOSE OF THE STUDY

The Purpose of this study was to introduce a new learning environment in which augmented reality technology is used for learning English, and to motivate the students belonging to rural or countryside area of China both intrinsically and extrinsically, which will affect their learning outcomes in learning English. To examine the students more, their behaviors, while using the AREWL and cognitive attainment were analyzed. To answer all the research questions shown in section 4.3.

## 4.3 RESEARCH QUESTIONS

Q1.  What are the achievements in English words learning of students using AREWL?

Q2.  Were there both intrinsic and extrinsic motivations involved in the learning process?

Q3. What is the relation of student's intrinsic and extrinsic motivation with their learning achievement?

Q4. How do students behave while using AREWL?

Q5. What w the cognitive attainment of students using AREWL?

Q6. What is the relation between behavioral patterns and cognitive attainment of students?

Q7. What are the opinions of teacher and students about the use of AREWL?

## 4.4 SELECTION OF SAMPLE AND POPULATION

This study was conducted in a Private elementary school named as Blueberry School in Liangxiang, which is the country side area of Beijing China. The population of Liangxiang county is about 112,000, it have administrative divisions, which contains around 40 administrative villages and 26 communities, According to [15] English level of students from rural areas are low as compared to the students in urban areas, the reason behind this lower level of English is that the schools mostly use old traditional ways of English teaching, and the lack of use of advance technologies., the ESL teachers are usually non-native, as compared to the other rural side schools of Beijing, who have hired a native teacher, as well as the use of technology,, is also obvious. We went to the Blueberry private elementary school, to discuss the research we wanted to conduct, after taking permission, we describedthe experiment process. Purposive sampling technique is used in this study because we choose the sample for a specific purpose. The sample consists of 5 non-native English teachers, who were teaching English to different grades of that elementary school, from grade 1 to grade 5, and the application designed in this study, have a teacher portal. Teacher ha been given several privileges, e.g., he/she can see the learning progress of the students and also can take a quiz in weekly lessons, using the application. The student sample was selected according to their knowledge about the content being taught in the lessons of the application, first a pretest was taken among 40 students of grade 1 and 2 of elementary schools, and 20 among them were selected for further experiment, the selection of those 20 students were based on their scores from pretest, those having lowest scores in the pretest were selected., so we divided students and teachers in a group, we gave 4 students to each teacher, who will assist students during the experiment and they also were the part of that experiment.

## 4.5 DATA COLLECTION TOOLS

In this study data collection, tools were mostly chosen according to the requirements of the research questions from the literature, except the pretest-posttest which was application based and the question asked in that were developed for testing the knowledge presented in the application for teaching words. Rest of the research tools are adopted from the literature.

We wanted to collect data regarding student's intrinsic and extrinsic motivation, student achievements behavioral patterns of students, cognitive attainment and opinion of students about AREWL, and teacher's opinion about AREWL. We will discuss all of them below.

### 4.5.1 Pretests and Posttests (AREWL Evaluation of student's achievement)

The electronic form of pretest and posttest were conducted to reveal the effectiveness of learning process regarding achievements. Each teacher first gave Tablets/Smartphones to the student to check if they already know about the content of the application. The teachers first choose 40 students to check their knowledge about the content of application, then among them 20 those students were selected for the experiment which got the lowest scores, so after the experiment, the teacher took a test again which was considered as the posttest using AREWL to find the learning achievements of the students. The scores from the test taken by the teacher using the learning application AREWL was used for analysis and finding the relation between the intrinsic and extrinsic motivation versus learning achievement. The picture below shows the score of students achieved in the quiz taken by the teacher using AREWL.

The questions in the pretest and posttest were identical. We tested three things,

1. For example, there is a name of Animal "Wolf"
2. First, the student have to recognize the animal
3. He/she have to Read to check the pronunciation
4. Choose the correct spelling

### 4.5.2 Questionnaire (Intrinsic and Extrinsic Motivation of students)

For finding students intrinsic and extrinsic motivation, 5points Likert scale method was used, and the statements for this survey was adopted from [17] with changes according to our research and also there was Chinese translation for the ease of the students, the researcher and teachers were helping the students to understand the meaning of the statement if they feel any difficulty. CronBach's Alpha for this study was 0.79.

Following statements were used to find students intrinsic and extrinsic motivation Q1 to 5 are for finding intrinsic motivation and from Q5 to Q8 for finding extrinsic motivation

1. I am interested in using AREWL app for English learning
2. I like to see the virtual objects in real word
3. I like to play with AREWL at home and learn new words
4. I enjoy Practicing spelling and pronounce learning after the interactive learning of AREWL
5. I am interested to unlock each lesson to see and play with new 3D objects daily of AREWL app.
6. I want to show good scores in class learning evaluation activity for AREWL app.
7. I want to learn English using AREWL to compete, my friends
8. I want to get the best English learner of the week award, using AREWL in weekly competition.

### 4.5.3 Video Recordings & Observation Forms (Students Behavioral Pattern)

For collecting data regarding student's behavioral pattern we prepared an observation form, which [114] used for their study to find students and teacher behavioral patterns, the form they used mostly had reading behaviors of children readings, parents narrating and interaction with the AR elements and the book, for this study we selected some major behaviors which were related to our study, i.e., students interaction with the content of application, interaction with AR elements, operating the AREWL and some other behaviors like distraction and inspecting. Videos were recoded to record all the observations of the students while they were playing with the application and those videos were used to note down the behaviors on the observation form. The videos were recorded on a mobile phone camera and data collected from those videos were analyzed after the experiment using observation forms. Cronbach's Alpha value for this study was 0.81 the table (3) below shows the behavior selected from [114] study.

Table 3: Data Collection tool from literature to analyze Behavioral patterns

| Data Collection Tool | Behavioral Categories | Behaviors | Description |
|---|---|---|---|
| Behavioral pattern ([114, 115]) | Children Behavior of operating AREWL | Control turning | The child control the operation of AREWL |
| | | | The child turn the camera of the mobile device to see the object from different angles |
| | Child's interaction-oriented behaviors regarding the content of the app | Pointing | The child points at the details of the content in App |
| | | Commenting | The child comments about the content of the app |
| | | Questioning | The child ask questions about the content of the app |
| | | Responding | The child respond to the questions asked in the app or by the teacher |
| | | Repeating | The child repeats the word after listening in the app, or after the teacher |
| | Children's interaction-oriented behaviors regarding the AR elements | Pointing | The child points at the details of AR elements |
| | | Commenting | The child makes a comment on the AR elements |
| | | Questioning | The child ask questions about AR element |
| | | Responding | The child respond to the teacher questions or comments or the AR elements |
| | | Repeating | The child repeats the teacher's additional information regarding the AR elements |
| | Others | Distraction | The child distracts during the process of spelling typing and word listening. |
| | | Inspecting | The child inspects the AR elements and tries to touch it |

## 4.5.4 Interview forms (Opinion and Cognitive attainment of students)

For student's opinion and cognitive attainment interview forms were used, for knowing the opinion of students about the application few simple questions about the application were asked, whereas for cognitive attainment [114] method was used, extensive description and appearance description were known by those questions in the interview form Table 4 shows the questions asked for knowing the opinion and cognitive attainment of the students.

Table 4: Interview Form for Student's Opinion & cognitive attainment

| **The opinion of Children**: |
| --- |
| 1. Do you like AREWL? Why? Can you explain it? <br> 2. Which was the most attractive feature for you? Why it is, can you tell me? |
| **Cognitive attainment [114]:** |
| 3. What elements of AR did you see on these flash cards? How do they look like can you describe them? Do they have any features which are distinguishing? <br> 4. What are your thoughts about the content presented by AREWL? For example, the animation you have watched, or any other details about spelling learning or the assessment part you want to share with us? |

## 4.5.5 Questionnaire (Teacher's opinion)

The opinion of teachers about the application w known by Likert scale of teacher acceptance model (TAM), we selected four factors from the survey of [116] as per our requirements are shown in Table 5 i.e. attitude towards using ARWL, perceived usefulness, behavioral intentions and perceived ease of use. For this we used 11 items and analysis of reliability were determined and we found the value of Cronbach's Alpha as 0.83

Table 5: Survey Form for getting teacher's opinion

| **Attitude toward using AREWL** |
| --- |
| 1. AREWL makes work more interesting. <br> 2. Working with AREWL is fun <br> 3. I look forward those aspects of my job that require me to use AWL |
| **Perceived usefulness** |
| 1. Using AREWL will improve my work.<br>2. Using AREWL will enhance my effectiveness.<br>3. Using AREWL will increase my productivity |
| **Behavioral intention to Use** |
| 1. I will use AREWL in the future<br>2. I plan to use AREWL often. |
| **Perceived ease of use** |
| 1. My interaction with AREWL is clear and understandable<br>2. I find it easy to get AREWL to do what I want it to do.<br>3. I find AREWL easy to use |

## 4.6 EXPERIMENT PROCESS

We organized the whole experiment in the break time of the students, the whole research process took one week.

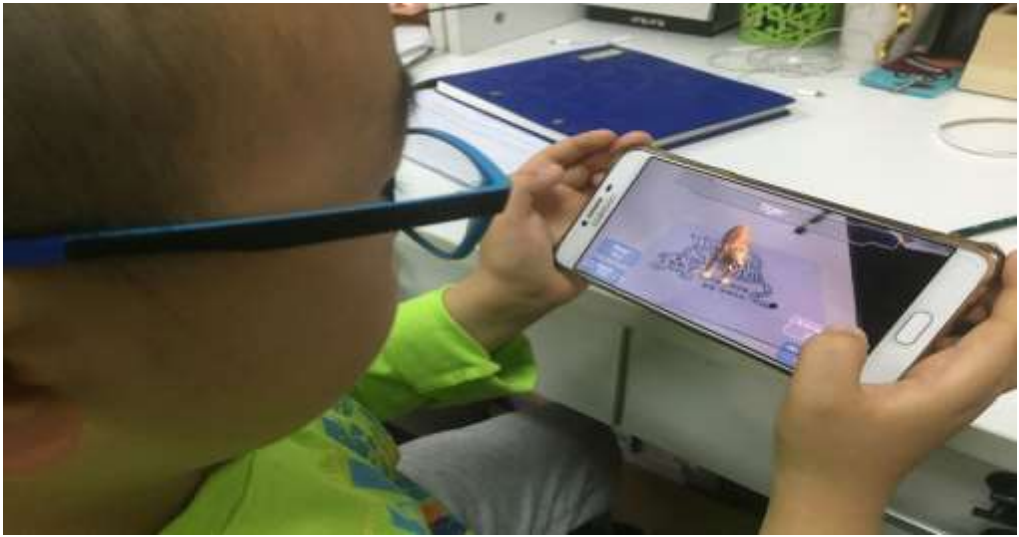

Figure 17: Student Using AREWL during the experiment

### 4.6.1 Day 1

We went to the school to discuss the experiment we wanted to do in that school, we described the whole research process to the administrator of the school, and five teachers willingly agreed with us to help us in the completion of the process. We requested the teachers to find 20 students among 40 students who have less knowledge about the contents taught in AREWL pretest scores who can bring phone or tablets to play the application. We with the help of school teachers requested the parents to arrange Android smartphones or tablets for this activity, 4, to 5 smartphones we arranged and about 17 students brought their devices with their parents' permission, so we divided the pretests groups into two. We arrange this activity in the break time of the students; the pretest session took 15 minutes for one group and 15 minutes for the second group.

### 4.6.2 Day 2

We assigned one teacher to 4 students, the students registered with that teacher using the application. Each student started with the first lesson named as "Animals" The teachers firstly explained to every student about the use of the application, every student had tablets or phone to play freely with the application, all the interaction with 3d objects, and spelling learning and then the assessment, while they were playing was recorded on video cams, for further data analysis of students behavioral patterns, the whole activity took 20 minutes, then we let the students who did not pass the assessment to play the application at home to pass the assessment to unlock the next lesson for the next day.

### 4.6.3 Day 3

On the third day, the experiment was the same, but learning lesson was different, this lesson was about learning fruit names, the whole process was the same as day 2, but this time the students were familiar with the application so was easy for them to use, mostly every student had tried the application in home as we were checking the teacher portal for the attempts and mistake of every student using teacher portal.

### 4.6.4  Day 4

This day we asked the teachers to take a quiz in the two lessons which the students already learned, so the teacher initiated the quiz, in which the questions were randomly generated for that two lesson which the students already learned. In this way, we collected the data which showed each students learning achievement ranking wise.

### 4.6.5  Day 5

We started collecting data from students and teachers, to answer our research questions. As there were many data collection tools and each one was useful for answering a specific question of the current study. Firstly we took teachers opinion on the survey form; then we distributed the questionnaires for finding intrinsic and extrinsic motivation, lastly for finding the cognitive attainment and opinions of students we used the tools mentioned in section 4.5.4.

## 4.7  DATA ANALYSIS PROCEDURES

We will discuss some data analysis methods used in this study; we used SPSS software to make the analysis process quicker and accurate. Three types of analysis were used,

- descriptive analysis,
- Correlational analysis
- Regression Analysis

In this section, we will discuss these types and will show its use in our study.

### 4.7.1  Descriptive Analysis

Descriptive analysis is one of the important types of analysis which helps to describe, summarize and show the data in such a way which is meaningful, for example, data might emerges some patterns. In other words, it is simply a way to describe the data, it is important because it is difficult to visualize data if it is in raw form, especially if the data is very big. So descriptive analysis allows simple data interpretation. Descriptive statistics of data can be presented in from of tables, charts, and graphs, or discussion of the results (i.e., statistical commentary) [117].

## 4.7.2 Correlational analysis

To find the relationship between two variables, correlational analysis used, which shows the strength of the relation between two variables. If the correlation between two variables high or strong, it shows that there is a strong relationship between those two variables. Similarly, if the correlation is weak or low, it means that the relationship between these two variables is low. Studying the strength of the relationship between two variables with the availability of statistical data is called correlational analysis, SPSS software is commonly used in this process, to fin the strength of the relationship between variables. A correlation coefficient is produced after all the statistical process which tell us the information. Pearson r is the most widely used correlational coefficient, from this kind of analysis it is assumed that the measurement of analysis of two variables w done on the least interval scales. Which means that the measurement is done on increasing value range. The correlational coefficient is calculated by dividing the covariance of variables by the product of their standard deviations. The formula for finding correlational coefficient is given below.

$$Y = a + bX$$
$$b = \frac{N\sum XY - (\sum X)(\sum Y)}{N\sum X^2 - (\sum X)^2}$$
$$a = \frac{\sum Y - b\sum X}{N}$$

Equation 1: Formula of Correlational Coefficient

Whereas 'r' is the correlational coefficient, n shows the number of sample and x and y are the two variables.

## 4.7.3 Regression analysis

The process of regression analysis is used to estimate the relationships between dependents and one or more independent variables. More specifically this type of analysis helps to understand the extent of change in the value of the dependent variable if there is a change in the value of any one of the independent variable. While the value of other independent variables is held fixed. The most common use of regression model is estimating the conditional expectation of the variable which is dependent, and the given independent variables, which in other words is the average value of the variable which is dependent when there is a fixed value of independent variables.

The wide use of regression model is for forecasting and prediction, whereas it also has used in the field of machine learning. It is also used for understanding that which variable among the independent variables have relation with the dependent variable, and also for the exploration of the forms of these relationships. If the circumstances are restricted, this type of analysis can be used to find casual relationships between dependent and independent variables. However, this may lead to false relations or illusions.

There are many techniques available, which are developed for regression analysis. Most common are ordinary least common and linear regression which are parametric, in this technique of regression analysis, the number of unknown parameters are finite and are estimated from the data, for which the function of regression is defined, On the other hand, nonparametric regression is a technique in which the regression function is allowed to lie in a set of function which is specified and which might be infinite-dimensional.

In this study, we used linear regression to find the relation between independent and dependent variables. The equation for linear regression is given below [118]

$$Y = a + bX$$

$$b = \frac{N\sum XY - (\sum X)(\sum Y)}{N\sum X^2 - (\sum X)^2} \quad \& \quad a = \frac{\sum Y - b\sum X}{N}$$

Equation 2: Formula for Regression Analysis

Where

N= number of observations, or years

X= a year index (decade)

Y= population size for given census years.

## 4.8 ANALYSIS PROCEDURE FOR EACH RESEARCH QUESTION

Analysis of the data was done using different techniques, to find the results of the above questions, Descriptive, correlational and regression analysis were done. In the next sections, we will point out the analysis methods used for analysis of data related to each question.

*Question 1(Students Achievement):* By the achievement of students here means the scores of posttest taken using AREL, students achievements were analyzed by descriptive analysis, for this quantitative method were used to find the mean and standard deviations of learning achievements of the students.

*Question 2 (Intrinsic and extrinsic motivations of students):* To find out if the students are motivated both intrinsically and extrinsically after using AREWL, descriptive analysis of 5 scales Likert survey was done using quantitative methods.

*Question 3 (Relations between Intrinsic, Extrinsic motivations and achievements):* For answering this questions, correlational analysis methods were used, following relations were revealed between these three variables

1. Relation between student intrinsic motivation and student achievements,
2. Relation between student extrinsic motivation and student achievements,
3. The relation between both intrinsic, extrinsic motivation with student achievements.

*Question 4 (behavioral patterns):* We used observation forms for observing how the students behave while using AREWL, for providing reliable results, in this process firstly the researchers observed all the students behaviors, noted each behavior in the form, check the results twice and did correction in the data. Descriptive analysis w used to analyze percentage, counts, and standard deviations.

*Question 5 (Student's cognitive attainment):* We determined the cognitive attainment of students by two interview questions, the questions were related to the things they see, and the thoughts and imagination come in their minds, so the researcher analyzed these questions in two categories.

1. Extensive description
2. Appearance description

For example the statement of students "I saw a big tiger, it has black and orange lines" was evaluated as appearance description category, because he/she described the appearance of what he/she saw, as the students described about size of the tiger and its color, so it was counted as two frequencies. Similarly the statement of students "it seems that these animals are walking in front of me" was evaluated as extensive description and the frequency was counted as 1.

*Question 6 (relation between student's behaviors and cognitive attainment):* It is important to know the relation between the behaviors of students while using AREWL and their cognitive attainment, correlational method were used to analyze the data.

*Question 7 (Student's and teacher's opinion):* We did descriptive analysis teachers opinion, standard deviations, mean and frequency were discussed, and content analysis for students Opinions analysis, so teachers Opinion were analyzed in quantitative way and students opinion were analyzed qualitativelyy.

We double check all the data analysis process and took guidance from technology experts as we selected our data collection tools from the literature, therefore pilot study for making data collection tools was not done, as those tool is chosen from the literature were reliable. Table (4) show the details information about the tools and methods used for this study.

Table 6: Tools selected from the literature and Methods used for analyzing the data

| Research Question | Data Analysis Tool | Data Analysis Methods |
|---|---|---|
| Students Achievements | AREWL Based Pretest & Posttest | Descriptive Analysis |
| Intrinsic and extrinsic motivations of students | Questionnaire [17] | Descriptive Analysis |
| Relations between Intrinsic, Extrinsic motivations and achievements | Questionnaire & AREWL Posttest Results | Correlation and Regression Analysis |
| behavioral patterns of students | Videos and Observation Form [114], [115]) | Descriptive Analysis |
| Cognitive Attainment of the Students | Interview forms ([114], [115]) | Descriptive Analysis |

| Relation Between Cognitive attainment and behaviors | Interview form & Observation Form [114], [115] | Correlational Analysis |
|---|---|---|
| Teacher's Opinion | TAM survey ([116], [115] | Descriptive Analysis |
| Student's Opinion | Interview Form [115] | Content Analysis |

## 4.9 SUMMARY

In Chapter three we discussed the outline of our methodology for our research and the ways which we perform the whole research process, e.g. the design of our study, data collection methods and the tools used for the analysis of all the research questions, we also included that how we choose our sample for this study and how the tools for the collection of our data was chosen.

# 5 RESULTS AND ANALYSIS

## 5.1 ANALYSIS OF PRETEST AND POSTTESTS (STUDENT'S ACHIEVEMENT)

The combine group quiz was taken using AREWL by teacher, to let the students compete for each other after learning two lessons, (Animals and Fruits), We used the pretest and posttest results for three purposes, firstly the pretest was used to choose the sample group, secondly posttest and protests results were compared to check the learning effectiveness, at last we used the posttest results to find the relation of students posttest scores (achievements) with students motivations, which will be discussed in next section). The analysis of pretest (M=19.41, SD=8.07) and posttest (M=66.17, SD=17.81) shows the improvements in students' scores after the experiment. The mean score of the students before the experiment was below than 20 and after the experiment it reached above 65, which shows that learning English words with AREWL have positive learning effectiveness on students learning achievements.

Table 7: Descriptive analysis of pretest and posttest

|  | N | Mean | Std. Deviation |
|---|---|---|---|
| **Pretest** | 20 | 19.41 | 8.07 |
| **Posttest** | 20 | 66.17 | 17.81 |

## 5.2 ANALYSIS OF INTRINSIC AND EXTRINSIC MOTIVATION

One of the main point to explore in this study was to find if the students are motivated both intrinsically and extrinsically? , the data collected by the five scale Likert questionnaires taken from [17], were changed according to our requirement and analysis were done to find the mean and standard deviation. The results show some positive side of the students learning through AREL, the standard deviation and mean of both types of motivation are mentioned in the table (8). For intrinsic motivation (M=3.5, SD=0.64) shows that most of the answers of students was on the positive side, which indicates that students had strong intrinsic motivation, whereas extrinsic motivation (M=3.50, SD=0.59) whose mean and standard deviation values are a bit smaller than intrinsic motivation values, but still the standard deviation shows little variation and shows positivity, which also indicates that students had extrinsic motivation because of the weekly reward based quiz taken by the teacher.

Table 8: Descriptive Analysis of Intrinsic and Extrinsic Motivation

|  | N | Mean | Std. Deviation |
|---|---|---|---|
| **Intrinsic motivation** | 20 | 3.5176 | .64054 |
| **Extrinsic motivation** | 20 | 3.5098 | .59064 |

## 5.3 RELATION BETWEEN COMPONENTS OF MOTIVATION AND ACHIEVEMENT

In this study we discussed two types of motivations, Intrinsic and extrinsic, and its relation with learning achievement using AREWL, and to figure out if our system can make such environment in which the students are motivated intrinsically and extrinsically, so in the previous section we analyzed if intrinsic and extrinsic motivation exists or not, therefore in this section we showed the result of analysis being done to find the relation between these two types, intrinsic and extrinsic motivation with learning achievement of each student. So the results of correlation analysis between two variables (intrinsic motivation and achievement) and (extrinsic motivation and achievement) also result of regression analysis between two independent variables, Intrinsic and extrinsic motivation with the dependent variable result the following relations will be discussed below.

### 5.3.1 Correlation analysis between intrinsic motivation, extrinsic motivation, and achievement

In this study the correlation analysis between intrinsic motivation and achievements, extrinsic motivation and achievements were done by Pearson correlation analysis method. The results are shown in Table (9). All the variables (Intrinsic, extrinsic, and Achievements) were included in the analysis, so according to the results both intrinsic and extrinsic motivation had significance relation with the student learning achievements. The correlation between intrinsic motivation and achievement is (r=0.810, p<0.05) and the correlation between extrinsic motivation and achievement (r=.737, p<0.05) shows that there is a significance relation between these variables. Pearson correlation coefficient of intrinsic motivation and achievement is a bit greater than the correlation coefficient of extrinsic motivation and achievement, which shows that the linear relationship between intrinsic motivations with achievement is a bit stronger than the linear relation of extrinsic motivation with achievement. From this, we can conclude that intrinsic motivation has stronger relationship with student's English words learning achievement using AREWL.

Table 9: Results of Correlation Matrix of Components of Motivation and Achievements

|  | **Intrinsic motivation** | **Extrinsic motivation** | **Achievements** |
|---|---|---|---|
| **Intrinsic motivation** | 1 | | |
| **Extrinsic motivation** | .422*** | 1 | |
| **Achievements** | .810** | .737** | 1 |

**p < 0.01, *p < 0.05.

## 5.3.2 Regression Analysis between components of Motivation and Achievement

To further explore the relation between intrinsic motivation, extrinsic motivation, and achievement, we further did the analysis using regression method in which the relation between independent variables of motivation and the dependent variable of achievement were analyzed. Table (10) shows the result of the regression analysis between the independent and dependent variables. In this case dependent variable student learning achievement is not shown in the table, but its relation with two independent variables of motivation is shown. For the relation between intrinsic motivation and achievement (T-statistics= 5.52, p=0.00) T is greater than the constant value 2.55 and p < 0.05 shows that there is a significance relation between intrinsic motivation and achievement So which shows if a student is more intrinsically motivated using the AREWL he/she can have high achievement in their grades.On the other had the relation between extrinsic motivation and achievement (T= 2.85, P=0.04) shows that there is a small difference of between the constant and the value of T, but the T value is still greater, and p <0.05 show that there is somehow very small significance. But still the significance exist, which shows that effect of extrinsic motivation on students learning have not much significance.

The relation between both independent variables, intrinsic and extrinsic motivation together with dependent variable (T=4.58, p<0.05) shows that the T value is greater than 2.55. Which is the value of constant and the probability p<0.05 shows a strong relation between these variable, which means that if there are both kind of motivations intrinsic and extrinsic involved in the learning process, the learning outcomes will be more higher. The coefficient of determination R-Square shows that there 68.7 % of the variation in achievement is caused by the intrinsic and extrinsic and also together by the combine relation of two independent variables intrinsic and extrinsic motivation with achievement.

Table 10: Regression Analysis of Components of Motivation and Achievements.

|  | B | Std. Error | t-statistic | Prob. |
|---|---|---|---|---|
| **Constant** | -72.622 | 28.445 | 2.553 | .023 |
| **Intrinsic motivation** | 25.276 | 4.586 | 5.512 | .000 |
| **Extrinsic motivation** | 14.213 | 4.973 | 2.858 | .04 |
| **Both Intrinsic and Extrinsic Motivation** | 20.517 | 4.473 | 4.587 | .000 |
| **R-square** | .687 | | | |
| **F-statistics [Prob.]** | 15.363 [0.000] | | | |

## 5.4 ANALYSIS OF STUDENTS BEHAVIORAL PATTERNS

In this study, we did the analysis of Chinese student's behavioral patterns to explore how the students behave while using a new technology augmented reality for learning English words. As shown in the methodology part, we used [114, 115] method to analyze the behaviors and used observation forms, Table 11 shows the results of analysis of students behaviors. The codes counts, standard deviation, and percentage of occurrence of every behavior is mentioned. In total we counted 3895 codes, we divided the count codes in three major behavior category, named as SOMT, SICA, and SIAR. With SOMT we mean the behaviors of controlling, inspecting and turning the AREWL, SICA the behaviors noted while students were engaged in playing with the other contents, like spelling learning, evaluation, etc and SIAR shows those behaviors which students were showing while playing with the AR elements for learning. From the results, we can see that students pointed at the details of AR elements mostly (Count= 389, 9.98%). Moreover, the second highest count of the behavior was pointed at the other contents of AREWL (Count=387, 9.93%), Repeated after listening to the voice overs (Count=350, 8.89%) and asked questions about the daily lesson evaluation (count=333, 8.54%). In addition the students also comment and questions about AR elements (count=292, 7.49%) and (Count= 291, 7.49%). From the results, we can also see that students operating behaviors of AREWL (count= 777, 19.94%) were comparatively lower than students interaction oriented behaviors with other learning contents of AREWL (Count= 1663, 41.92%) and AR elements (Count= 1485, 38.12%).

Table 11: Students behavioral count of codes (N=20)

| | Behaviors Descriptions | Count | SD | Percentage |
|---|---|---|---|---|
| **Behaviors** | SOMT (Controlling, turning, inspecting)) | 777 | 4.9 | 19.94% |
| | SICA (Pointing, commenting, questioning, responding, repeating) | 1663 | 6.42 | 41.92% |
| | SIAR (Pointing, commenting, questioning, responding, repeating) | 1485 | 4.94 | 38.12% |
| | **Total Behaviors** | **3895** | **43.4** | **100%** |
| | Controlling | 226 | 4.9 | 5.80% |
| | Inspecting | 283 | 6.42 | 7.21% |
| | Turning | 268 | 4.94 | 6.80% |
| | Pointing (CA) | 387 | 4.57 | 9.93% |
| | commenting (CA) | 251 | 2.87 | 6.44% |
| | Questioning (CA) | 333 | 4.01 | 8.54% |
| | Responding (CA) | 310 | 4.85 | 7.95% |
| | Repeating (CA) | 350 | 4.76 | 8.98% |
| | Pointing (AR) | 389 | 3.23 | 9.98% |
| | commenting (AR) | 292 | 4.84 | 7.49% |
| | Questioning (AR) | 291 | 3.73 | 7.47% |
| | Responding (AR) | 268 | 2.68 | 6.88% |
| | Repeating (AR) | 247 | 3.26 | 6.34% |
| | **Total Behaviors** | **3895** | **43.4** | **100%** |

SOMT: Students Behavior of operating AREWL
SICA: Student's Interaction- oriented Behavior regarding other Content of AREWL
SIAR: Student's Interaction- oriented Behavior regarding AR elements
CA: Content of the AREWL
AR: AR Elements

## 5.5 ANALYSIS OF STUDENT'S COGNITIVE ATTAINMENT DURING AREWL ACTIVITY

In this study, we also determined the student's cognitive attainment while they were playing with the AREWL, as we followed [114, 115] methods to analyze cognitive attainment, in our study we in total found 256 codes, which were revealed regarding appearance description and extensive description. From the results it is shown that in total 194 codes were counted as appearance description category, and 62 codes were counted as extensive description category, which refers to the imagination of students while playing ARWEL activity, which after analysis is explained that student's appearance description (M=9.7, SD=3.96) is quite better than extensive description (M=3.1, SD=1.55). All these results are shown in Table 12.

Table 12: Count codes for students cognitive attainment (N=20)

| Cognitive Attainment | Total | Mean | SD | Max | Min |
|---|---|---|---|---|---|
| Appearance Description | 194 | 9.7 | 3.96166 | 18 | 3 |
| Extensive Description | 62 | 3.1 | 1.55259 | 7 | 1 |

## 5.6 THE RELATION BETWEEN STUDENT'S BEHAVIORS AND COGNITIVE ATTAINMENT

We examined the relation between the behaviors of students and their cognitive attainment, as its one of the important factor to analyzed the relations between these two, previous researchers who researched on teaching using augmented reality applications, analyzed the behavioral patterns and cognitive attainment and also the relation between these two variables [114, 115]. So we also considered it as an essential part of the analysis of these two variables on Chinese leaners. We used Pearson correlation analysis to find the correlation, and the results are shown in Table (13). We included student's learning behaviors and three of their categories (SOMT, SIAR, and SICA) and also the two categories of cognitive attainment, appearance, and extensive description. From the results, we found out that student's behavior had relation with both extensive and appearance description. It was revealed that the behaviors of Pointing CA (r=0.209, p<0.05), Pointing AR (r=0.49, p<0.05) and Commenting AR (r=0.52, p<0.05) had relation with appearance description, while Turning AREWL (r=0.31, p<0.05), Questioning CA (r=0.21, p<0.05), and Questioning AR (r=0.097, p<0.05) had a significance relation the extensive description. Lastly, we can say from the results that SOMT and SIAR had relationships with appearance description and SICA had a relationship with an extensive description.

Table 13: The correlation between students behavior and cognitive attainment (N=20)

|  | **Appearance description** | **Extensive description** |
|---|---|---|
| **COMT** | **0.831*** | 0.967 |
| **SICA** | 0.28 | **0.629*** |
| **SIAR** | **0.906**** | 0.788 |
| Controlling | 0.41 | 0.86 |
| Inspecting | 0.315 | -0.54 |
| Turning | 0.562 | **0.314*** |
| Pointing (CA) | **0.209*** | 0.63 |
| commenting (CA) | **.472*** | -0.251 |
| Questioning (CA) | 0.41 | **0.211**** |
| Responding (CA) | 0.49 | -0.284 |
| Repeating (CA) | 0.88 | 0.36 |
| Pointing (AR) | **0.49*** | 0.46 |
| Commenting (AR) | **0.52*** | -0.093 |
| Questioning (AR) | 0.54 | **0.097*** |
| Responding (AR) | 0.64 | 0.255 |
| Repeating (AR) | 0.12 | 0.42 |

**$**p < 0.01$, $*p < 0.05$**

SOMT: Students Behavior of operating AREWL

SICA: Student's Interaction- oriented Behavior regarding other Content of AREWL

SIAR: Student's Interaction- oriented Behavior regarding AR elements

CA: Content of the AREWL

AR: AR Elements

## 5.7 GRAPHICAL REPRESENTATION OF STUDENTS DAILY LESSON BASED EVALUATION

In this study, we also had student daily lesson based evaluation, after learning each lesson there was a self-evaluation. The main purpose of this was to test the student's learning of daily lesson, the student can attempt as many times as he/she wants, and the count of attempts and the mistake during the last attempts were updated in the teacher's portal. So teacher has the idea of student's learning progress. The two graph ha has shown below shows the number of attempts and mistakes done by each student while practicing lesson 1, and 2. Graph 1 and Graph 2.

The orange lines in the graph show a number of mistakes in the last attempt and blue lines shows the number of attempts of each lesson assessment. From the bar lines of this graph it is clear that students who had practiced more, their mistakes count was getting lower, in other words, we can say that if attempts are more than mistakes are less and if mistakes are more it means that lesson was not practiced much.

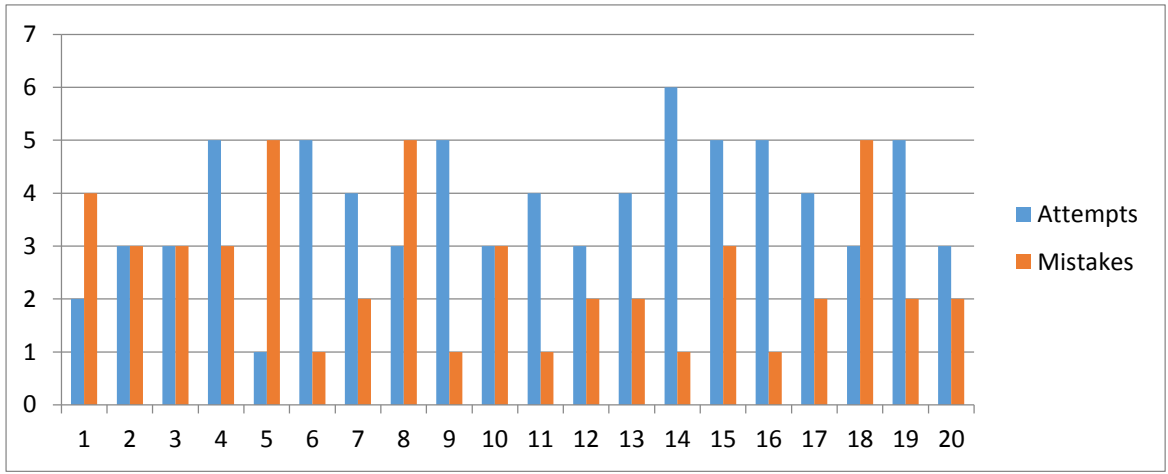

Graph 1: Lesson 1 attempts & Mistakes (N=20)

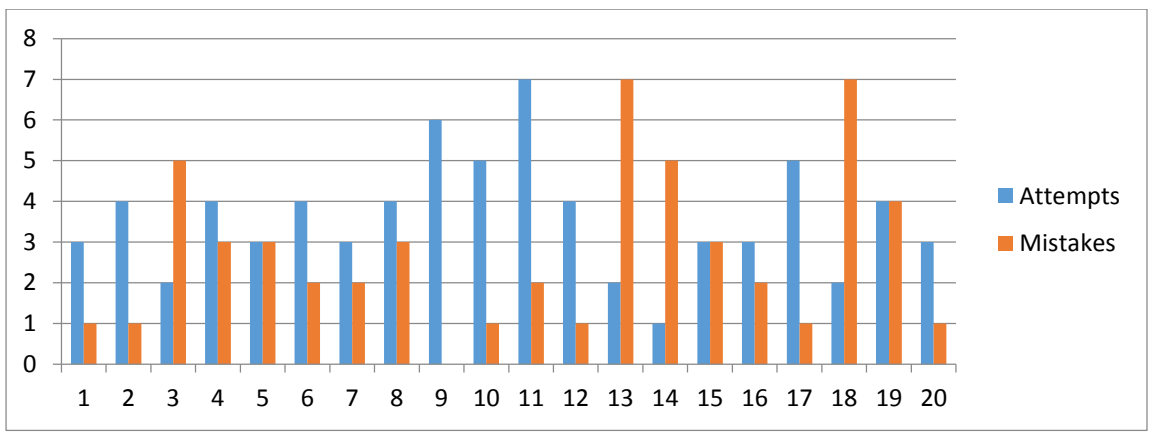

Graph 2: Lesson 2 attempts & Mistakes (N=20)

## 5.8 ANALYSIS OF TEACHERS OPINION ABOUT AREWL

In this study, we determined the teacher's opinion about the AREWL activity to know about the acceptance level of school teachers for the implementation of this English words learning application. There were total five teachers involved in the experiment, technology acceptance model (TAM) used by [115] was adopted, and analysis was done in the term of perceived usefulness, perceived ease of use, attitude towards using AREWL and behavioral intention to use. Questions related to those four points were asked to find out if they have positive or negative attitude towards this learning activity, from the results, we found out that all the teachers like

AREWL activity (f=5), which shows how much the teachers found this activity useful to be implemented in future, and they said that students of grade 1 and grade 2 were both enjoying to interact with 3d objects, and were willing to pass the daily lessons. After examining the opinions in details, we found that the value of perceived usefulness was high (M=3.1333, SD=0.73030). The general attitude of teachers for this activity was positive (M=3.0083, SD=0.55388), the analysis of ease of use shows a lower mean and a high value of Standard Deviation (M=2.9000, SD=0.65192), which means some of the teachers thought that the application is a bit difficult to use and some of them answered very positively, that's why there is a high variation, which is clear from its SD. Their attitude towards using AREWL was positive, (M=3.0000, SD=0.33333) and also they showed positive behavior for future use of AREWL (M=3.0000, SD=0.50000). The details of teacher's opinion analysis are shown in Table 14.

Table 14: Results of Teachers Opinion

| **Constructs** | N | Mean | Std. Deviation |
|---|---|---|---|
| Perceived usefulness | 5 | 3.1333 | 0.73030 |
| Perceived ease of use | 5 | 2.9000 | 0.65192 |
| Attitude toward using EM | 5 | 3.0000 | .33333 |
| Behavioral intention to Use | 5 | 3.0000 | .50000 |
| **Total** |  | **3.0083** | **0.55388** |

## 5.9 ANALYSIS OF STUDENT'S OPINION ABOUT AREWL

We examined the opinion of students about learning English words using AREWL, All the students like AREWL activity f=20, some of them like animations, while others like the magical view of objects, they found the spelling learning and interactive learning both and enjoyed learning using AREWL, they further told us that it's funny and interesting, Fig 14 shows the detailed explanation of the content analysis of students opinion about AREWL.

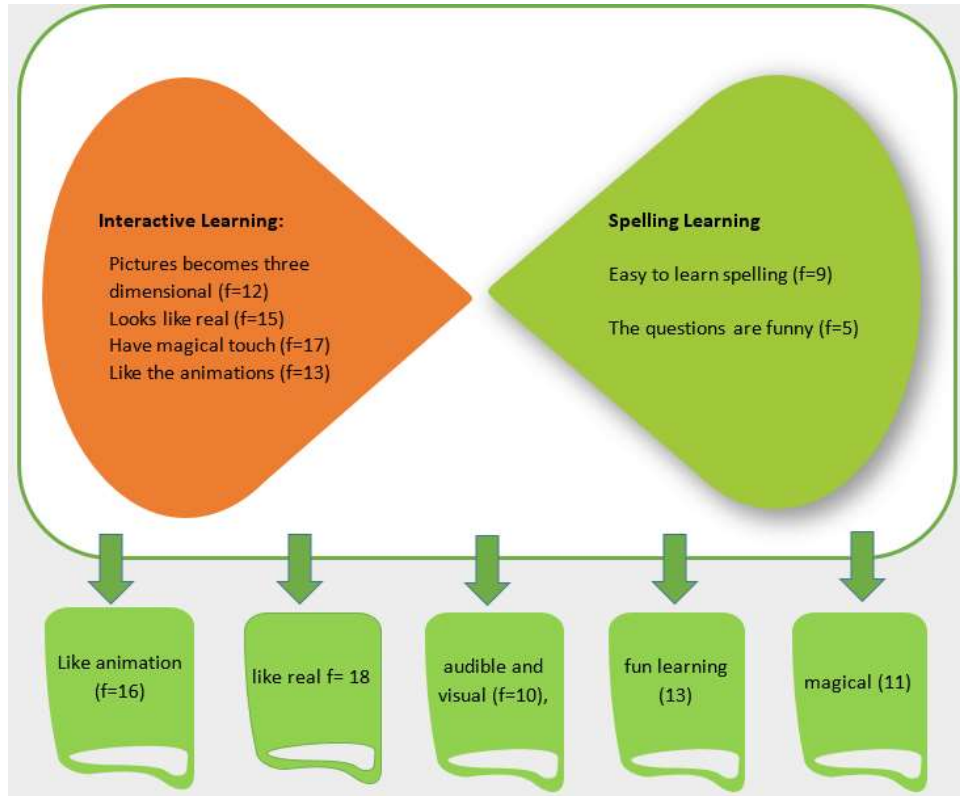

Figure 16: Content analysis of student's opinion

## 5.10 SUMMARY

In this chapter, we show all methods used for doing the analysis and also the results of our study. We did the analysis of all the collected data, the analysis methods included, descriptive analysis, correlational analysis, and regression analysis. The results from our analysis are mostly positive and are near to our expectations, from the results we can see that both the components of motivation have strong and significance relation with student learning achievement, in the design of application we considered emphasizing on the values of intrinsic and extrinsic motivation, which affected the learners learning achievements in terms of scores. Further, we did an analysis of behavioral pattern and cognitive attainment, to find out the way of interaction with the technology of elementary school students, and the analysis of teachers and student's opinion about AREWL shows us that both students and the teachers were interested in the activity and they liked it.

# 6 SUMMARY AND CONCLUSION

## 6.1 SUMMARY OF THE STUDY

In this study we tried to answer several questions related to the problems of low motivation in Chinese students from rural areas. In the first chapter we described the problems and the proposed solutions, after thorough study of literature we found out that the main reason of low achievements in English learning among rural area students is that they are not motivated enough because of many factors which include quality resources of education. So our approach is to let those students take an interest in English learning even if they are living in not so much developed environment and included teachers and schools in our experiment because that is the place and the people who have the main role in students motivations towards any learning process. From the literature review, we also came to know that such environments are essential for learning which create both intrinsic and extrinsic motivation in students toward learning of a specific subject [17], for which our study is English. So we decided to choose rural or country side area because it is mentioned in the previous researchers there is an education Gap between rural and urban people level of education. There is a need of implementation of cheap technology which can motivate students to learn with interest. Old traditional ways of teaching, which are followed in rural area schools, is also one of the main cause of low motivation. In our study we wanted to know if intrinsic and extrinsic motivations are emphasized equally in the learning environment of AREWL, if yes then what could be its impact on learning achievements, so from the results of this experiment, we found that using ARWEL gave both intrinsic and extrinsic motivations to the students, and their relations with learning achievements were proved from the Correlation and Regression analysis, which showed that intrinsically motivated students had somehow better score than extrinsically motivated students, but those students who were both motivated intrinsically and extrinsically using AREWL had the highest scores among all.

In the previous studies researchers who used augmented reality for learning purpose examined the behavior and cognitive attainment of the students [17, 114, 115], as these factors are important to examine because the Augmented reality is mostly interactive way of leaning. So the time when they are playing, their behavior towards the learning contents and attainment and imaginations which come into their minds should be examined. Our examination of behavioral patterns shows the behaviors of pointing to the AR elements and pointing to the other contents, e.g. spelling learning activity and daily lesson based evaluation questions. Also, repeating, commenting and questionings was also among the behaviors which the students show more. Which are not similar to [114, 115] study. In the study of [114] the behavior shown by the students was mostly controlling, and in [115] study the students mostly prefer pointing, responding, inspecting, and turning. In learning experience behavior is one of the most important features [119]. In some cases, if the interaction in learning activity is poor because of poor interactive design, results in cognitive overload and stress [115, 120, 121]. Therefore Interaction is considered important feature in the learning process.

The results of student's cognitive attainment show that there was a bit low cognitive attainment in the learning process. We calculated 62 codes in extensive appearance category and 194 from the appearance description. The results are somehow higher than [114, 115] studies, because in our study there were evaluation based on every lesson learning, if we notice in our behavioral patterns results among the most prominent behaviors were pointing, questioning, repeating and responding. While they were learning interactively, they pointed to the AR elements and asked questions, in spellings learning they were repeating the alphabets, and words, while solving the daily lesson based evaluation they asked questions, and also they respond to the teacher's questions.

This study also revealed the relation between Student's Behavior and Cognitive attainment in which SIAR (Pointing, commenting, questioning, responding, repeating for AR elements) behaviors of pointing) and SOMT (Controlling, turning, inspecting) were related to appearance description, that is why they were describing more what they saw the appearance of the objects and was controlling and turning AREWL to inspect the appearance of the 3d objects. Also. SICA(Pointing, commenting, questioning, responding and repeating behaviors related to other Contents of AREWL ) was related to the extensive description. That is because there were spelling, learning and evaluation parts in each lesson, so they asked more questions, respond to the teacher, AREWL and repeated the words taught by the AREWL. Which shows that students were more active in learning Activity of AREWL.

Lastly, we did the analysis of teacher's and student's opinion about AREWL which was positive in all cases, both teacher and student's like AREWL. Teacher opinion is important in introducing the new technology because the teacher will be using it together with students in future.

## 6.2  CONCLUSION & FUTURE WORK

From this study we conclude that both intrinsic and extrinsic motivations are important features for learning outcomes of students, technology like Augmented reality if used for learning can bring change in learners motivation and achievement, as it is important to implement it in undeveloped areas where technology use is rare and students and teachers are unaware of the use of this new technology, due to which learning gap in rural and urban people are high. As Augmented reality is very cheap and not costly, as it is used on a mobile phone, which is very common these days. So poor people who do not have much access to expensive technology can integrate this technology into learning because from our experiment we considered the opinions of students and teacher about this technology and they had positive answers which show that they like the augmented reality way of learning.

**Future Work**:

As every study is not completed, and there is always some ways of improvement. From our thinking, we suggest some future improvements to the current system. As follow.

1. To enhance the magical perspective of AR learning Smart Glasses can be used in classrooms to give a more attractive and fun learning environment.
2. Voice Recognition can be added to make the pronunciation learning more advance.
3. Quality multimedia content should be added with a pedagogical approach for the development of a proper interactive English learning application.
4. Teacher portals can be more advanced than the one developed for our Demo Application. In which teacher and students are more close to each, other regarding communication.
5. Different samples can be used for testing AR English learning application in different countries.

# Acknowledgments

I would like to thank and praise the all-knowing *ALLAH* to impart me little knowledge from unlimited treasures of knowledge. Furthermore, I would like to thank Prof. Zhendong Niu for supervising my master's research and for his great interest throughout my studies. His invaluable guidance, encouragement, and understanding have been a great source of motivation for me. I am deeply obliged to him for helping significantly to my doctoral studies.

I wish to express my gratitude to the academic staff and management of our school, my lab mates and batch fellows for always helping and guiding me to get through the difficult times. I also want to thank China Scholarship Council (CSC) for their funding for my master's studies. This would not have been possible for me without their support.

Also, I am thankful to all my friends and well-wishers. Their affections have encouraged me and brought more delight to the quality times than I ever could have dreamt. I want to thank for their moral support during my studies. I also thank to them for motivating and guiding me though this profound challenge.

Finally, sincere thanks to my father, mother, wife, loving son and siblings for their prayers, patience, source of inspirations and support throughout my studies at BIT.